\documentclass[journal]{IEEEtran}
\usepackage{geometry}
\usepackage{booktabs}

\usepackage{sidecap}
\usepackage{graphicx}
\usepackage{colortbl}
\usepackage{xcolor}
\usepackage[linesnumbered,ruled,vlined]{algorithm2e}
\usepackage{xcolor}
\usepackage{soul}
\usepackage[most]{tcolorbox}
\tcbuselibrary{skins, breakable}

\newtcolorbox{keyfinding}[1][]{
  enhanced,
  breakable,
  colback=teal!5!white,
  colframe=teal!70!black,
  coltitle=white,
  fonttitle=\bfseries\sffamily,
  title={\ding{72} Key Finding},
  left=6pt, right=6pt, top=4pt, bottom=4pt,
  boxrule=1pt,
  arc=3pt,
  shadow={1mm}{-1mm}{0mm}{black!20},
  #1
}

\newtcolorbox{threatalert}[1][]{
  enhanced,
  breakable,
  colback=red!4!white,
  colframe=red!70!black,
  coltitle=white,
  fonttitle=\bfseries\sffamily,
  title={\ding{43} Threat Alert},
  left=6pt, right=6pt, top=4pt, bottom=4pt,
  boxrule=1pt,
  arc=3pt,
  shadow={1mm}{-1mm}{0mm}{black!20},
  #1
}

\newtcolorbox{recommendation}[1][]{
  enhanced,
  breakable,
  colback=blue!4!white,
  colframe=blue!60!black,
  coltitle=white,
  fonttitle=\bfseries\sffamily,
  title={\ding{52} Recommendation},
  left=6pt, right=6pt, top=4pt, bottom=4pt,
  boxrule=1pt,
  arc=3pt,
  shadow={1mm}{-1mm}{0mm}{black!20},
  #1
}

\newtcolorbox{insightbox}[2][]{
  enhanced,
  breakable,
  colback=yaleblue!5!white,
  colframe=yaleblue,
  coltitle=white,
  fonttitle=\bfseries\sffamily,
  title={#2},
  left=6pt, right=6pt, top=4pt, bottom=4pt,
  boxrule=1.2pt,
  arc=3pt,
  shadow={1mm}{-1mm}{0mm}{black!20},
  #1
}

\newtcolorbox{researchgap}[1][]{
  enhanced,
  breakable,
  colback=orange!5!white,
  colframe=orange!70!black,
  coltitle=white,
  fonttitle=\bfseries\sffamily,
  title={\ding{73} Research Gap},
  left=6pt, right=6pt, top=4pt, bottom=4pt,
  boxrule=1pt,
  arc=3pt,
  shadow={1mm}{-1mm}{0mm}{black!20},
  #1
}

\newtcolorbox{definitionbox}[2][]{
  enhanced,
  colback=gray!5!white,
  colframe=gray!60!black,
  coltitle=white,
  fonttitle=\bfseries\sffamily,
  title={#2},
  left=6pt, right=6pt, top=4pt, bottom=4pt,
  boxrule=0.8pt,
  arc=3pt,
  #1
}
\usepackage{enumitem}
\usepackage{tcolorbox}

\definecolor{yaleblue}{RGB}{0, 53, 107}

\definecolor{red}{RGB}{255, 0, 0}
\definecolor{llmcolor}{RGB}{70,130,180}

\usepackage{algorithm2e}

\usepackage{amsmath}

\SetCommentSty{mycommfont}
\SetKwInput{KwInput}{Input}              
\SetKwInput{KwOutput}{Output}            
\usepackage{floatrow}
\usepackage{listings}
\usepackage{wrapfig}
\usepackage{float}
\usepackage[hidelinks]{hyperref}

\usepackage{amssymb}
\usepackage{pgfplots}
\pgfplotsset{width=10cm,compat=1.9}
\usepackage{graphicx}
\usepackage{lipsum}
\usepackage{tikz}
\usepackage{tabularx, booktabs}

\usepackage{multirow}
\usetikzlibrary{patterns}
\usepackage{textcomp}
\usepackage[utf8]{inputenc}
\usepackage{newunicodechar}
\newunicodechar{≤}{\ensuremath{\leq}}
\pgfplotsset{compat=1.8}
\usepgfplotslibrary{statistics}
\usepackage{tabularx, booktabs, xcolor, pifont}

\usepackage{xspace}
\usepackage{lineno,hyperref}
\usepackage[utf8]{inputenc}
\usepackage{geometry}
\usepackage{multirow}
\usepackage{array}
\usepackage{longtable}
\usepackage{algpseudocode}
\usepackage{amsmath}
\usepackage{comment}
\usepackage{cite}

\ifCLASSINFOpdf

\else
 
\fi

\begin{document}

\newgeometry{left=0.8in,top=0.8in,bottom=0.8in}
\title{Securing Cryptography in the Age of Quantum Computing and AI: Threats, Implementations, and Strategic Response}

\author{
    \IEEEauthorblockN{Viraaji Mothukuri, Reza M. Parizi}\\
     \IEEEauthorblockA{Decentralized Science Lab, College of Computing and Software Engineering, Kennesaw State University, GA, USA
    \\ \ \ vmothuku@students.kennesaw.edu, rparizi1@kennesaw.edu}
 }

\maketitle

\begin{abstract}
This review examines how quantum computing and artificial intelligence challenge current cryptographic systems. We analyze the literature to assess the resilience of algorithms against quantum attacks (Shor's and Grover's algorithms) and AI-enhanced cryptanalysis. RSA and elliptic curve cryptography are at risk of compromise from quantum computers. Symmetric algorithms like AES-128 retain security, but with a reduced effective key length under quantum attacks. Deep learning models demonstrate improved side-channel analysis, extracting keys from protected implementations. These convergent threats require a defense-in-depth approach that combines post-quantum algorithms, implementation hardening, and cryptographic agility. We find that lattice-based algorithms (ML-KEM, ML-DSA) resist known quantum attacks but require careful implementation to prevent side-channel leakage. Hash-based signatures (SLH-DSA) provide conservative security with signature sizes ranging from 17 to 50 KB. No single approach addresses both quantum and AI threats comprehensively. Organizations must treat cryptographic security as an ongoing process rather than a fixed deployment, maintaining the capability to update algorithms as threats evolve.
 \end{abstract}

\section{Introduction} \label{sec:intro}

In contemporary digital infrastructure, cryptographic security provides the foundational layer for authentication, confidentiality, and data integrity. The Bank for International Settlements reports that global foreign exchange trading alone reached \$9.6 trillion per day in April 2025, with over-the-counter interest rate derivatives adding a further \$7.9 trillion per day \cite{bis2025triennial}. These figures capture only a subset of electronically secured financial activity. When securities settlement, card payment networks, and interbank transfer systems such as Fedwire and SWIFT are included, the aggregate volume of cryptographically protected financial transactions substantially exceeds \$10 trillion daily. Beyond finance, modern systems maintain healthcare records requiring decades of protection, secure classified government communications, and protect industrial intellectual property worth trillions. Each of these domains depends on the computational intractability of specific number-theoretic problems to guarantee the confidentiality and integrity of data in transit and at rest. The sheer scale of this dependency means that any disruption to the underlying mathematical hardness assumptions would propagate across virtually every sector of the digital economy simultaneously. This concentration of risk in a small number of cryptographic primitives motivates the urgency of the threat analysis presented in this work.

Traditional cryptographic systems rely on well-established mathematical hardness assumptions that have provided stability for decades. RSA encryption depends on the computational difficulty of factoring large composite integers \cite{rivest1978method} and continues to protect the majority of encrypted communications worldwide nearly five decades after its introduction. Elliptic curve cryptography achieves equivalent security with substantially smaller key sizes, where a 256-bit ECC key provides security comparable to a 3072-bit RSA key \cite{koblitz1987elliptic, miller1986use}, enabling efficient deployment across TLS, SSH, and resource-constrained environments. Diffie-Hellman key exchange relies on the discrete logarithm problem in finite fields and remains central to session key establishment in virtually all modern protocols \cite{diffie1976new}. These constructions share a critical property in that they appear computationally infeasible using classical algorithms but admit polynomial-time solutions on sufficiently powerful quantum computers.

However, two emerging technological disruptions now challenge these foundational assumptions through distinct attack vectors. Quantum computing targets the mathematical algorithm layer through polynomial-time solutions to factorization and discrete logarithm problems, rendering the entire class of deployed public-key cryptographic systems vulnerable to a single algorithmic advance. Artificial intelligence targets the physical implementation layer through neural network-based side-channel analysis, extracting cryptographic keys from power consumption traces, electromagnetic emanations, and timing variations during cryptographic operations. The convergence of these threats creates exceptional complexity for defenders, requiring organizations to simultaneously address mathematical vulnerabilities in cryptographic algorithms and physical vulnerabilities in their implementations. The Advanced Encryption Standard, selected by NIST in 2001 \cite{daemen2002design}, has withstood two decades of intensive cryptanalysis without practical attacks on full-round implementations, though its symmetric-key structure places it in a fundamentally different 
 \restoregeometry
vulnerability category under quantum attack. Grover's algorithm reduces the effective security of AES-128 from 128 bits to 64 bits, representing a meaningful degradation but not a catastrophic one, whereas Shor's algorithm \cite{shor1994algorithms} reduces the effective security of RSA and ECC from their nominal classical levels to zero.

\subsection{Quantum Computing Threat Landscape}
\begin{threatalert}[title={\ding{43} Quantum Threat Acceleration}]
Recent algorithmic work reduced RSA-2048 factorization requirements by \textbf{95\%} (from 20M to $<$1M physical qubits). Combined with vendor roadmaps converging on 100,000+ physical qubits by the early 2030s and error correction ratios of 250:1 to 1000:1, the timeline for cryptographically relevant quantum computers has shortened significantly. Expert probability estimates have \textbf{doubled} since 2022.
\end{threatalert}

Quantum computing has progressed from a theoretical possibility to an active engineering challenge, with recent hardware milestones indicating that the threat to deployed cryptographic systems is no longer distant but approaching within institutional planning horizons. The trajectory of progress across multiple hardware platforms suggests that the question facing organizations is not whether cryptographically relevant quantum computers will emerge, but whether current migration timelines are sufficient to complete the transition before they do. This framing is consequential because it shifts the burden of proof. Organizations that defer migration must justify the assumption that quantum computing timelines will extend beyond their data sensitivity periods, rather than migration advocates needing to prove that the threat is certain. The following paragraphs examine recent hardware breakthroughs, algorithmic improvements in quantum resource estimation, and expert probability assessments that collectively inform the current understanding of threat timelines. Each line of evidence contributes independently to the conclusion that cryptographic migration planning should begin for organizations with data sensitivity periods exceeding ten years. The evidence is organized by hardware demonstrations, algorithmic advances, and expert consensus to enable readers to assess the weight of each component separately. Considerable uncertainty remains in all three areas, and we note limitations and caveats throughout.

Google's Willow chip, announced in December 2024, achieved below-threshold error correction with 105 superconducting qubits \cite{google2024willow}. This milestone demonstrates that errors decrease exponentially as physical qubit count increases, validating the fundamental path to fault-tolerant quantum computing. Willow completed a calculation requiring $10^{25}$ years classically in five minutes, though this benchmark problem has no cryptographic relevance. The significance lies in validating the error correction scaling behavior rather than demonstrating cryptanalytic capability. Prior to this result, a persistent concern in the field was that adding physical qubits to a system would increase total error rates faster than error correction could suppress them, effectively preventing the construction of large-scale fault-tolerant machines. The Willow results demonstrate that this concern is surmountable in superconducting architectures under specific conditions. However, the gap between 105 physical qubits and the approximately one million physical qubits estimated for RSA-2048 factorization remains substantial, spanning roughly four orders of magnitude.

Additional hardware progress across alternative architectures demonstrates that superconducting qubits are not the sole path to fault tolerance. A neutral atom tweezer array containing 6,100 highly coherent atomic qubits with coherence times of 12.6 seconds has been demonstrated \cite{Manetsch_2025}, establishing a new scale for neutral atom platforms. Single-qubit gate error rates of $1.5 \times 10^{-7}$ (0.000015\%) have been achieved using microwave-controlled trapped calcium ions operated at room temperature without magnetic shielding \cite{smith2025singlequbit}, representing the lowest gate error recorded to date and reducing the overhead required for quantum error correction. Two-qubit gate fidelities exceeding 99.99\% have been demonstrated using trapped-ion technology without ground-state cooling \cite{hughes2025trappediontwoqubitgates9999}, addressing the complementary challenge of entangling gate accuracy. These results span three distinct hardware platforms, namely superconducting circuits, neutral atoms, and trapped ions, indicating that progress toward fault-tolerant quantum computing is not contingent on any single technological approach. Multiple vendor roadmaps now converge on 100,000 or more physical qubits by the early 2030s \cite{ibm2023roadmap}, though the translation from physical to logical qubits depends on error correction overhead ratios that currently range from approximately 250:1 to 1000:1 depending on the error correction code and target logical error rate. At these ratios, 100,000 physical qubits would yield on the order of 100 to 400 logical qubits, which remains below the threshold required for cryptanalytic applications but represents substantial progress toward that threshold.

Recent algorithmic advancements have substantially reduced quantum resource requirements for cryptanalytic applications. A 95\% reduction in physical qubit requirements for RSA-2048 factorization has been demonstrated, decreasing estimates from approximately 20 million to under 1 million physical qubits \cite{gidney2025quantum}. This improvement derives from two primary optimizations in the form of approximate residue arithmetic and enhanced surface code implementations. The result assumes error rates of $10^{-3}$ per gate and the use of specific error-correction codes, meaning that different assumptions yield different resource estimates and introduce uncertainty about the actual requirements. Hybrid quantum-classical algorithms enable NISQ devices to contribute to cryptanalysis before achieving full fault tolerance \cite{wang2024hybrid}. Variational quantum eigensolvers and quantum approximate optimization algorithms tackle subproblems while classical computers handle the remainder, creating graduated threat levels rather than a binary distinction between vulnerable and secure states. Consequently, quantum computers that reach stated vendor targets would have sufficient physical qubit counts to approach the threshold for breaking current public-key cryptography, assuming error-correction overhead remains manageable. The resource estimates represent current understanding, not guaranteed future capabilities, and unforeseen engineering challenges could significantly extend timelines, while conversely, algorithmic breakthroughs could accelerate them.

Expert assessments of quantum threat timelines have accelerated correspondingly. The Global Risk Institute's 2024 survey of 47 quantum computing experts indicates that the probability of cryptographically relevant quantum computers has risen to 34\% by 2034, with estimates doubling from the 2022 assessment, which projected 17\% by the same date \cite{mosca2024quantum}. The acceleration stems from the combination of recent breakthroughs in error correction, revised vendor roadmaps, and algorithmic improvements reducing resource requirements by an order of magnitude. However, expert forecasts in quantum computing carry substantial uncertainty. The field has faced repeated delays over the past two decades, and unforeseen engineering challenges in areas such as interconnect scaling, cryogenic infrastructure, and control electronics could further extend timelines beyond current projections. The probabilities represent consensus estimates based on current progress rather than certainties about future capabilities. Despite this uncertainty, the trend shows consistent acceleration, with probability estimates rising uniformly across successive survey years rather than oscillating.

\subsection{Artificial Intelligence Threat Landscape}

Artificial intelligence presents a more nuanced threat landscape than popular discourse suggests. Technical analysis reveals a sharp divergence between AI capabilities across different attack surfaces. Neural networks achieve high success rates in attacking cryptographic implementations through side-channel analysis, extracting keys from power consumption traces and electromagnetic emanations with efficiency that surpasses classical statistical methods. Large language models fail comprehensively at algorithm-level cryptanalysis, achieving zero percent success on any properly implemented modern encryption algorithm across all tested models. This bifurcation reflects fundamental architectural constraints inherent to current neural network designs rather than temporary limitations that will be overcome by scaling model size or training data. Understanding this distinction is essential for accurate threat assessment because conflating neural network side-channel capabilities with general AI reasoning about cryptographic algorithms can lead to either overestimation or underestimation of the actual threat, depending on which capability is generalized. The following subsections examine each capability class in detail, documenting both the evidence and the theoretical basis for the observed divergence.

\subsubsection{Neural Network Capabilities}

Neural networks demonstrate significant effectiveness in side-channel analysis because the task involves pattern recognition in continuous analog signals, a domain where convolutional architectures demonstrate well-established strengths. Deep learning models have achieved single-trace key recovery against protected AES implementations \cite{Kim_Picek_Heuser_Bhasin_Hanjalic_2019}, extracting 128-bit keys from a single power consumption measurement, whereas classical differential power analysis requires millions of traces for equivalent key recovery. Cross-device attack accuracy exceeding 99.9\% has been achieved with approximately 10$\times$ lower Minimum Traces to Disclosure compared to classical correlation power analysis \cite{xdeepsca2019}, demonstrating that models trained on one physical device can successfully attack different devices implementing the same algorithm. Recent work demonstrates key extraction from electromagnetic emanations at distances of 25 meters under laboratory conditions using consumer hardware \cite{cagli2024electromagnetic}, extending attack range beyond traditional side-channel distances, though signal quality degrades with distance, and real-world environments introduce noise absent in controlled settings. Transfer learning enables attacks on device variants with minimal per-device training, reducing the data requirements that previously limited the practical applicability of profiled side-channel attacks.

The threat extends to post-quantum cryptography implementations. Single-trace power analysis attacks against CRYSTALS-Kyber key generation were published at TCHES 2025 \cite{hal05455454}, demonstrating that newly standardized algorithms are vulnerable to the same class of implementation attacks that affect classical cryptography. Deep learning attacks have successfully defeated Rotating S-boxes Masking, a protection scheme specifically designed to resist conventional power analysis, achieving substantial reductions in the number of required traces compared to classical methods \cite{kuroda2021rsm}.

\subsubsection{Large Language Model Limitations}

Large language models fail comprehensively at algorithm-level cryptanalysis. The CipherBank benchmark evaluated 18 state-of-the-art language models on 2,358 decryption problems spanning nine cipher algorithms \cite{jiang2025cipherbank}. Claude 3.5 Sonnet achieved the highest accuracy at 45.14\%, with OpenAI o1 achieving 40.59\%. Even the highest-performing models failed on the majority of polyalphabetic ciphers developed in the early modern period, and every tested model achieved exactly 0\% success on properly implemented modern encryption algorithms, including AES, RSA, and elliptic curve cryptography. These results indicate that decades of cryptographic engineering have successfully eliminated the statistical patterns that neural networks exploit in simpler ciphers.

Mechanistic interpretability research documents the architectural basis for these failures. Causal intervention studies demonstrated that large language models employ collections of memorized heuristics rather than algorithmic reasoning, with 91\% of arithmetically important neurons classified into distinct heuristic types that implement pattern-matching rather than algorithmic steps \cite{nikankin2024arithmetic}. The GSM-Symbolic study showed 12--15\% performance variance across mathematically equivalent problems, confirming reliance on surface patterns rather than abstract mathematical understanding \cite{mirzadeh2024gsm}. Communication complexity theory has been applied to prove that transformer architectures are mathematically incapable of composing functions when domains exceed relatively small sizes \cite{peng2024limitations}. Cryptographic algorithms involve multi-round compositional transformations over keyspaces of $2^{128}$ or larger, representing a fundamental architectural constraint rather than a limitation that additional training data or model scale will overcome.

\subsection{Harvest Now, Decrypt Later}

The harvest now, decrypt later threat model establishes migration urgency regardless of quantum timeline uncertainty \cite{harvestn}. Adversaries with strategic foresight can capture encrypted communications today and store them indefinitely for decryption once quantum computers become available. This attack requires no current cryptanalytic capability, only a network position enabling traffic interception and sufficient storage capacity. Both requirements are satisfied by nation-state actors and increasingly accessible to well-resourced criminal organizations. The temporal asymmetry of HNDL fundamentally distinguishes it from traditional threat models, where vulnerability and exploitation capability coincide. Data encrypted today using vulnerable algorithms becomes adversarially stored upon transmission, creating present exposure whose exploitation is deferred until quantum decryption capability becomes available. The Federal Reserve's September 2025 analysis highlights irreversible exposure in distributed systems, noting that blockchain technologies create permanent public records of encrypted transactions that remain vulnerable even after future algorithmic migrations.

Documented incidents consistent with HNDL operations illustrate the practical feasibility of large-scale traffic interception. In April 2020, Rostelecom announced BGP routes for over 8,000 prefixes belonging to major US technology companies including Google, Amazon, Facebook, and Cloudflare, rerouting traffic through Russian telecommunications infrastructure for approximately one hour \cite{thousandeyes2020rostelecom}. In June 2019, a BGP route leak from the Swiss data center operator Safe Host caused over 70,000 routes serving major European mobile networks, including Swisscom, KPN, Bouygues Telecom, and Numericable-SFR, to be redirected through China Telecom's network for more than two hours \cite{madory2019bgpleak}. While BGP route leaks can result from configuration errors rather than deliberate interception, the duration and scale of these incidents, combined with the involvement of state-affiliated telecommunications operators, has prompted sustained concern from the security research community. Joint guidance from CISA, NSA, and NIST explicitly warns that adversaries may be conducting harvest now, decrypt later operations against critical infrastructure, urging organizations to begin preparing quantum-readiness roadmaps \cite{cisa2023quantum}. The cautious phrasing of this guidance reflects intelligence sensitivity regarding the extent of ongoing collection activities, though policy responses from the United States, the European Union, and allied governments uniformly treat HNDL as an active threat requiring countermeasures rather than a hypothetical future concern. The convergence of documented BGP security failures with government warnings about strategic data collection creates a threat environment in which the technical prerequisites for HNDL operations are demonstrably satisfied by multiple nation-state actors. Organizations that transmit long-lived sensitive data over public networks must therefore assume that such data is subject to adversarial capture regardless of whether direct evidence of interception exists for their specific communications.

Section~\ref{sec:threat_model} formalizes the migration decision calculus through Mosca's risk equation, demonstrating that for organizations with long data sensitivity periods, urgency is determined by the sum of sensitivity and migration time rather than by the quantum threat timeline alone.

\subsection{Research Questions and Literature Gaps}

Existing literature treats quantum computing and artificial intelligence as separate threat domains addressed by distinct research communities. Post-quantum cryptography research focuses on algorithm design and standardization without systematic analysis of deployment complexity or implementation security challenges. AI cryptanalysis literature divides between side-channel analysis and algorithm-level attacks without distinguishing the fundamentally different capability profiles of neural networks and large language models. Deployment literature documents protocol-level integration without synthesizing progress across the full technology stack from hardware security modules through application-layer cryptographic libraries. Regulatory and policy analysis documents government mandates without technical depth regarding implementation feasibility or realistic migration timelines. This fragmentation creates operational risks because organizations may deploy post-quantum algorithms assuming complete protection while remaining vulnerable to side-channel attacks, or policymakers may mandate timelines disconnected from technical feasibility.

To address these gaps, this review is structured around four research questions that guide the analysis across the fragmented threat landscape.

\begin{itemize}
    \item \textbf{RQ1.} How do quantum computing and artificial intelligence create convergent yet distinct threats to cryptographic systems, and what are the fundamental architectural reasons for the divergence between neural network side-channel capabilities and large language model cryptanalytic limitations?
    \item \textbf{RQ2.} What is the actual state of post-quantum cryptography deployment beyond headline adoption statistics, and where do critical gaps exist between available standards and real-world implementation across different infrastructure layers?
    \item \textbf{RQ3.} What implementation security vulnerabilities affect newly standardized post-quantum algorithms, and how do the effectiveness profiles of existing side-channel countermeasures differ between classical and AI-enhanced attack methods?
    \item \textbf{RQ4.} What defense-in-depth framework is required to address threats that target fundamentally different layers of the cryptographic system stack, and how should organizations prioritize migration given sector-specific constraints and regulatory timelines?
\end{itemize}

These questions are motivated by the observation that no existing review synthesizes across quantum threat assessment, AI capability analysis, deployment measurement, implementation security evaluation, and migration planning into an integrated analytical framework. The fragmentation of the literature means that readers seeking guidance on post-quantum migration must consult separate bodies of work that employ different assumptions, threat models, and evaluation criteria. This review addresses post-quantum cryptographic migration as the primary domain, treating quantum computing and artificial intelligence as threat motivators that establish migration urgency rather than as subjects of a comprehensive technical survey in their own right, and aims to bridge those boundaries by providing analysis that is technically grounded in each domain while maintaining coherence across them.

\subsection{Contributions}

This work provides a comprehensive analysis of convergent threats from quantum computing and artificial intelligence to cryptographic systems. The contributions can be summarized as follows.

\begin{itemize}
    \item Providing dual threat synthesis analyzing both algorithm-level attacks via quantum computers and implementation-level attacks via neural networks, documenting the fundamental architectural reasons for AI capability bifurcation across attack surfaces and the implications of this bifurcation for defensive strategy
    \item Evaluating deployment reality beyond headline statistics, documenting critical gaps including 3.7\% server-side adoption versus 52\% client-side key exchange, near-zero digital signature migration, and decade-long timelines for critical infrastructure sectors
    \item Examining implementation security vulnerabilities in NIST's newly standardized post-quantum algorithms, including power consumption leakage in ML-KEM decapsulation and single-bit randomness vulnerabilities in ML-DSA signing, with a countermeasure taxonomy that maps the effectiveness of specific defenses against both classical and AI-enhanced attack methods
    \item Presenting a defense-in-depth framework mapping specific defenses to threat vectors across five system layers, demonstrating why no single defensive approach provides comprehensive protection against convergent threats and providing a risk assessment decision framework that integrates Mosca's equation with sector-specific migration constraints
\end{itemize}

The remainder of this paper is organized as follows. Section~\ref{sec:threat_model} presents the threat model analysis, examining the dual-threat attack surface, harvest-now-decrypt-later dynamics, and layered defensive requirements. Section~\ref{sec:litreview} provides the literature review spanning quantum computing capabilities, post-quantum standards and deployment, artificial intelligence in cryptanalysis, quantum key distribution, homomorphic encryption, quantum random number generation, hybrid protocols, public key infrastructure, hardware security, blockchain security, and testing and validation. Section~\ref{sec:analysis} presents the analysis, including threat timeline assessment, algorithm security evaluation, deployment metrics, the risk assessment decision framework, and research gaps. Section~\ref{sec:conclusion} concludes the paper.

\section{Threat Model Analysis}
\label{sec:threat_model}

This section formalizes the threat landscape introduced in Section~\ref{sec:intro} into three complementary analytical frameworks. These frameworks address the dual-threat attack surface model, the temporal dynamics of harvest-now-decrypt-later operations, and the layered defensive requirements these threats impose. Where the introduction establishes the evidence base for quantum and AI threats individually, this section analyzes how these threats interact structurally and what that interaction demands of defensive architectures.

\subsection{Dual-Threat Attack Surface}

\begin{figure}[ht!]
\centering
\includegraphics[width=\textwidth]{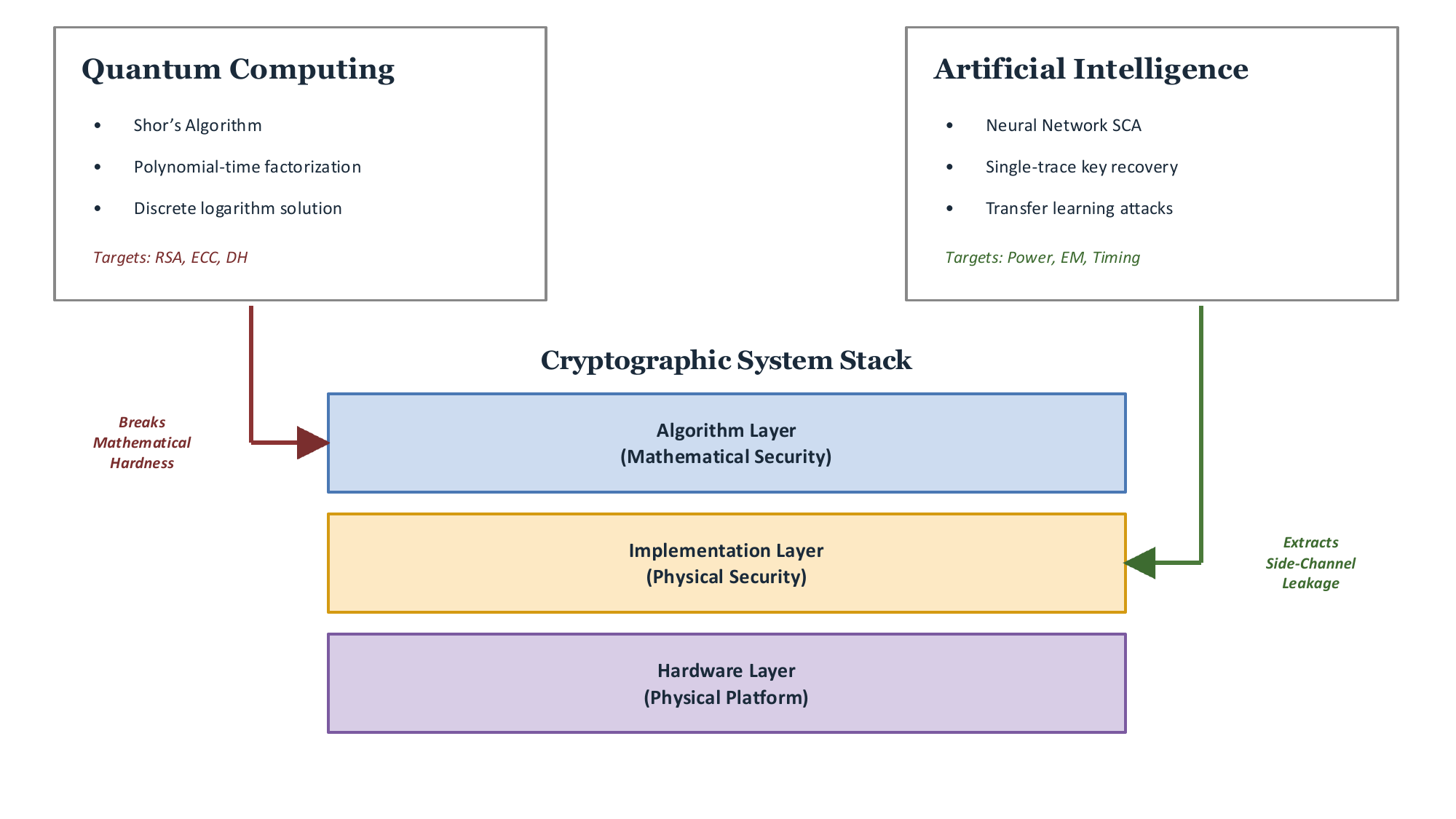}
\caption{Dual-threat attack surface model. }
\label{fig:dual_threat}
\end{figure}

Figure~\ref{fig:dual_threat} models the orthogonal relationship between quantum and AI threats to cryptographic systems. The central analytical observation is that these threat classes target fundamentally distinct layers of the cryptographic stack, creating attack vectors that cannot be addressed by any single defensive approach.

Quantum computing operates at the mathematical abstraction level. Shor's algorithm \cite{shor1994algorithms} renders RSA, ECC, and Diffie-Hellman key exchange insecure regardless of implementation quality by providing polynomial-time solutions to the underlying number-theoretic problems. As discussed in Section~\ref{sec:intro}, current hardware remains below the threshold of cryptanalytic relevance, but algorithmic improvements have reduced the estimated physical qubit requirement for RSA-2048 factorization by 95\% \cite{gidney2025quantum}, and expert probability estimates for cryptographically relevant quantum computers have doubled since 2022 \cite{mosca2024quantum}. A sufficiently capable quantum computer would break these primitives regardless of key length, padding scheme, or protocol context, rendering the threat fundamentally algorithmic in nature.

AI-based attacks operate at the physical implementation level. As documented in Section~\ref{sec:intro}, neural networks achieve single-trace key recovery against both classical and post-quantum implementations by recognizing data-dependent patterns in analog side-channel signals \cite{Kim_Picek_Heuser_Bhasin_Hanjalic_2019, hal05455454}. The critical analytical property of this vulnerability class is that it is algorithm-agnostic at the implementation layer: any algorithm executed on accessible hardware is potentially vulnerable regardless of its mathematical security properties, rendering the threat fundamentally physical rather than mathematical.

The orthogonality of these attack vectors carries direct implications for defensive strategy. Deploying post-quantum algorithms neutralizes quantum threats to the algorithm layer but provides zero protection against neural network side-channel extraction. Conversely, implementation hardening through masking, shuffling, and constant-time coding mitigates AI-enhanced physical attacks but leaves mathematical vulnerabilities to Shor's algorithm entirely unaddressed. An organization that completes post-quantum migration without implementation hardening has addressed one threat class while remaining fully exposed to the other. An organization that hardens its classical implementations without algorithm migration has the inverse exposure. The threat model therefore requires simultaneous mitigation across both layers, a requirement that Section~\ref{sec:layered_defense} formalizes into a five-layer defensive framework.

\subsection{Harvest Now, Decrypt Later Dynamics}

\begin{figure}[ht!]
\centering
\includegraphics[width=\textwidth]{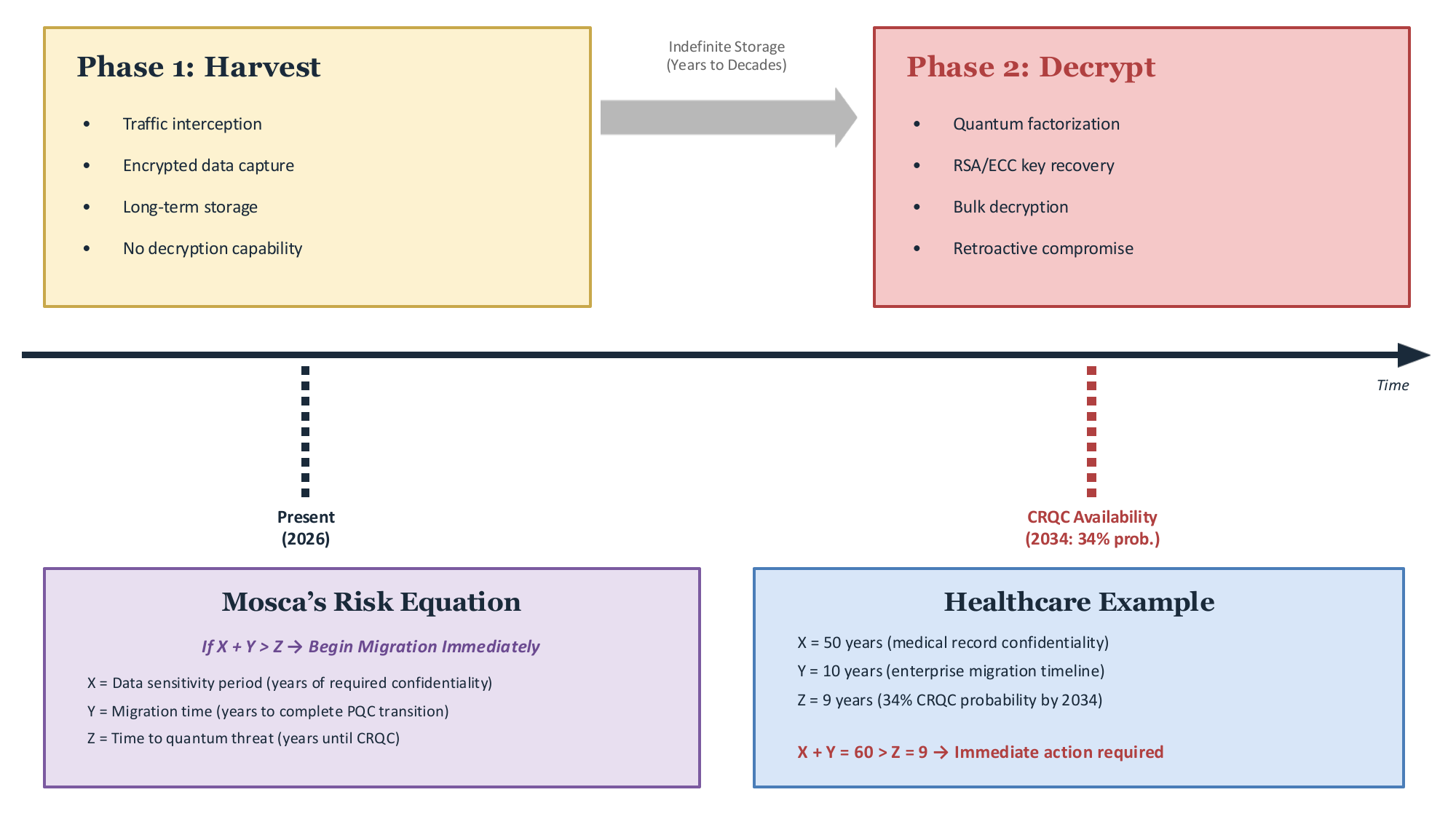}
\caption{Harvest now, decrypt later threat model.}
\label{fig:hndl}
\end{figure}

Figure~\ref{fig:hndl} presents the temporal structure of HNDL operations. Unlike traditional threat models where vulnerability and exploitation capability must coincide in time, HNDL decouples these requirements into three distinct phases, creating a threat that exists in the present tense regardless of when quantum decryption becomes feasible \cite{harvestn}.

During the \textit{harvest phase}, adversaries capture encrypted communications through network interception, BGP route manipulation, or direct access to communication infrastructure \cite{thousandeyes2020rostelecom, madory2019bgpleak}. This phase requires no cryptanalytic capability whatsoever. The only prerequisites are network position and storage capacity, both of which nation-state actors satisfy trivially and which well-resourced criminal organizations increasingly possess. Joint CISA, NSA, and NIST guidance explicitly warns that adversaries may be conducting HNDL operations against critical infrastructure \cite{cisa2023quantum}.

During the \textit{storage phase}, adversaries maintain captured ciphertext archives indefinitely. Storage costs decrease continuously according to well-established trends, while data value may persist for decades depending on the sensitivity of the underlying communications. This creates favorable economics for patient adversaries with strategic interests, as the cost of maintaining archives approaches zero over time while the potential intelligence value remains constant or appreciates as geopolitical contexts evolve.

During the \textit{decrypt phase}, which commences upon CRQC availability, quantum factorization enables bulk decryption of archived RSA and ECC-protected communications. Previously confidential data becomes accessible retroactively, with compromise severity determined by original data sensitivity rather than current relevance. The Federal Reserve's September 2025 analysis highlights that blockchain technologies create a particularly acute variant of this exposure, as permanent public records of encrypted transactions remain vulnerable regardless of future algorithm migration.

\subsection{Layered Defensive Requirements}
\label{sec:layered_defense}

\begin{figure}[ht!]
\centering
\includegraphics[width=\textwidth]{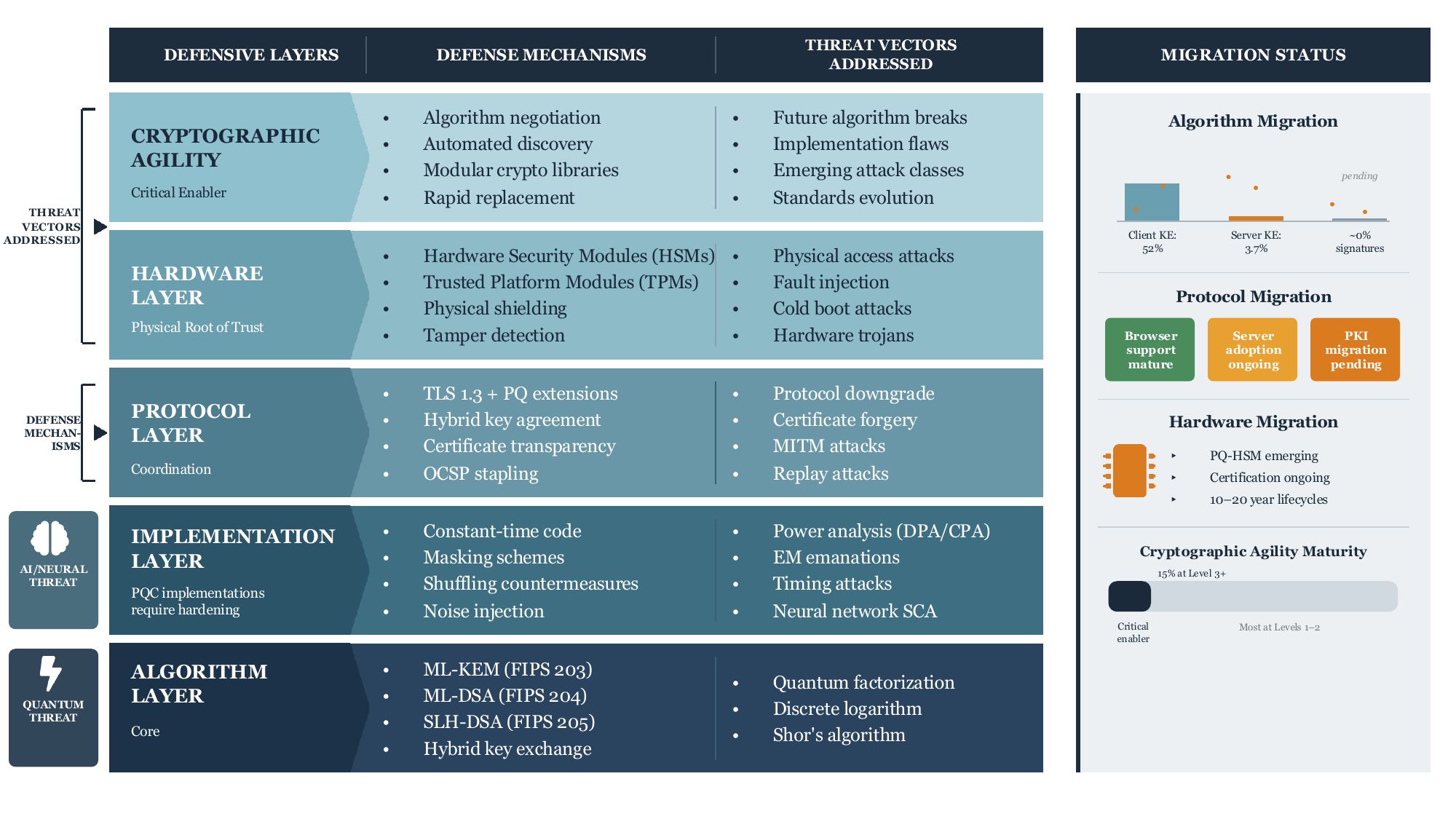}
\caption{Layered defensive requirements mapping countermeasures to threat vectors.}
\label{fig:defense_depth}
\end{figure}

Figure~\ref{fig:defense_depth} maps the defensive requirements arising from convergent threats across five system layers. The analysis demonstrates that comprehensive protection requires simultaneous deployment of countermeasures addressing distinct attack surfaces, and that current deployment metrics reveal dangerous asymmetries in migration progress across these layers.

The \textit{algorithm layer} addresses quantum threats through post-quantum cryptographic primitives. NIST-standardized algorithms (ML-KEM, ML-DSA, SLH-DSA) replace vulnerable number-theoretic constructions with lattice-based and hash-based alternatives \cite{nist2024pqc}. Deployment metrics reveal sharply asymmetric progress. Currently, 52\% of human web traffic utilizes hybrid post-quantum key exchange at the client side, driven by browser vendor defaults, while server-side adoption remains at 3.7\% \cite{cloudflare2024pq} and digital signature migration approaches zero. This disparity reflects fundamental differences in update mechanisms. Browser vendors push automatic updates to billions of clients simultaneously, whereas server operators must individually evaluate, test, and deploy algorithm changes across heterogeneous infrastructure. The near-zero signature migration is particularly consequential because digital signatures protect code integrity, firmware authenticity, and certificate chains, all of which are targets for retroactive forgery once quantum computers can break the underlying signature schemes.

The \textit{implementation layer} addresses AI-enhanced side-channel threats through physical countermeasures. Constant-time implementations prevent timing attacks. Masking schemes distribute sensitive values across multiple shares, requiring higher-order attacks that scale exponentially in the number of shares. Shuffling randomizes operation ordering to complicate trace alignment. Noise injection degrades the signal-to-noise ratio in side-channel measurements. Post-quantum implementations require dedicated hardening efforts because lattice-based algorithms exhibit vulnerability profiles distinct from classical cryptography, including leakage during rejection sampling, polynomial multiplication, and discrete Gaussian sampling \cite{hal05455454, primas2024masking}. The gap between algorithm standardization and implementation hardening creates a window during which organizations may deploy mathematically secure algorithms in physically vulnerable configurations.

The \textit{protocol layer} provides transport security through TLS 1.3 extensions supporting hybrid key agreement. Browser vendor implementation has matured, with major browsers defaulting to post-quantum-enabled cipher suites. Server adoption proceeds more gradually due to infrastructure complexity and backward compatibility requirements. Public key infrastructure migration remains pending, requiring coordinated updates across root certificate authorities, intermediate CAs, certificate transparency logs, and OCSP responders \cite{sikeridis2024interop}. The PKI dependency chain means that protocol-layer migration cannot complete until the entire trust hierarchy supports post-quantum signatures, a coordination problem that extends timelines well beyond individual server upgrades.

The \textit{hardware layer} isolates cryptographic operations within tamper-resistant boundaries. Hardware security modules and trusted platform modules protect key material from physical extraction. Post-quantum HSM certification is ongoing, but equipment lifecycles spanning 10--20 years create extended migration timelines for embedded and critical infrastructure deployments \cite{nist2024cmvp}. Organizations cannot simply update firmware on deployed HSMs in many cases. Physical replacement of tamper-resistant hardware is required, introducing procurement, installation, and recertification costs that dominate the migration timeline for sectors such as banking, defense, and industrial control systems.

\textit{Cryptographic agility} enables adaptation to evolving threats through algorithm negotiation and modular library architectures \cite{nist2024agility}. Organizations achieving higher agility maturity levels can replace compromised algorithms within days rather than months. However, survey data indicate only 15\% of organizations reach Level 3 maturity or above \cite{gri2024maturity}, with the majority remaining at foundational inventory and assessment stages. This agility deficit compounds the risks identified above. Organizations that lack the architectural capacity to swap algorithms rapidly face compounded exposure if any deployed post-quantum algorithm is broken, as the cryptanalytic compromise of SIKE during the NIST evaluation process demonstrated \cite{nist2024pqc}.

The layered analysis reveals a structural asymmetry in how the two threat classes penetrate the defensive stack. Quantum computers penetrate the algorithm layer, bypassing all implementation, protocol, hardware, and agility defenses below it. Neural networks penetrate the implementation layer, bypassing algorithmic security regardless of whether classical or post-quantum primitives are deployed. These attack vectors operate independently, rendering any single-layer defense insufficient against the combined threat. Defense-in-depth across all five layers is therefore mandatory rather than optional for organizations facing both threat classes within their planning horizons.

\section{Literature Review} \label{sec:litreview}

\subsection{Quantum Computing Capabilities and Threats}

Google's Willow chip achieved below-threshold error correction with 105 superconducting qubits in December 2024, demonstrating that error rates decrease exponentially with increasing qubit count \cite{google2024willow}. Prior to this result, a persistent concern was that adding physical qubits to a system would increase total error rates faster than error correction could suppress them, effectively preventing the construction of large-scale fault-tolerant machines. The Willow result demonstrates that this concern is surmountable for superconducting architectures under specific conditions. Willow completed a calculation requiring $10^{25}$ years classically in five minutes, though this benchmark problem has no cryptographic relevance. The significance lies in validating the path to fault-tolerant quantum computing rather than demonstrating cryptanalytic capability.

IBM targets 100,000 qubits by 2033 through modular approaches connecting multiple quantum processors \cite{ibm2023roadmap}. Quantinuum's trapped-ion systems achieved a quantum volume of $2^{20}$ \cite{quantinuum2024volume}, demonstrating alternative architectural paths. Coherence times improved from 20 to 100 microseconds across platforms. However, these metrics do not directly measure cryptographic threat. Quantum volume abstracts architecture-specific details and does not indicate performance for factoring or discrete logarithm operations. Error correction overhead means 1 million physical qubits may yield only thousands of logical qubits available for computation.

On the algorithmic front, recent work reduced RSA-2048 factorization requirements from 20 million to under 1 million physical qubits using approximate residue arithmetic and optimized surface codes \cite{gidney2025quantum}. This reduction assumes error rates of $10^{-3}$ per gate and the use of specific error correction codes, meaning that different assumptions yield different resource estimates and introduce uncertainty about actual requirements. Hybrid quantum-classical algorithms enable NISQ devices to contribute to cryptanalysis before achieving full fault tolerance \cite{wang2024hybrid}. Variational quantum eigensolvers and quantum approximate optimization algorithms tackle subproblems while classical computers handle the remainder, creating graduated threat levels rather than a binary distinction between vulnerable and secure states.

The resource estimates represent current understanding, not guaranteed future capabilities. Unforeseen engineering challenges could significantly extend timelines, while conversely algorithmic breakthroughs could accelerate them. The gap between demonstrated capability and cryptographically relevant quantum computing remains significant but is narrowing, driven by concurrent hardware progress and algorithmic improvements across multiple platforms.

\subsection{Post-Quantum Cryptography Standards and Deployment}

NIST finalized three post-quantum cryptographic standards on August 13, 2024, concluding an eight-year evaluation process \cite{nist2024pqc}. ML-KEM (FIPS 203) provides lattice-based key encapsulation with public keys of 800--1568 bytes and ciphertexts of 768--1568 bytes, based on the hardness of the Module-LWE problem. ML-DSA (FIPS 204) addresses digital signatures ranging from 2420 to 4595 bytes, using Module-LWE and Module-SIS problems. SLH-DSA (FIPS 205) provides hash-based signatures ranging from 7856 to 49856 bytes, relying solely on collision-resistant hash functions. Security levels map to AES equivalents, with Level 1 approximating AES-128, Level 3 approximating AES-192, and Level 5 approximating AES-256, accounting for Grover's algorithm, which halves the effective security of symmetric keys.

The standardization process revealed the difficulty of assessing post-quantum security with confidence. SIKE (isogeny-based) and Rainbow (multivariate) were cryptanalytically compromised during evaluation, demonstrating that the absence of known attacks differs fundamentally from proof of hardness. Lattice-based cryptography relies on problems such as the Shortest Vector Problem and the Learning with Errors problem, with worst-case-to-average-case reductions providing a theoretical foundation. However, ML-KEM and ML-DSA utilize module lattices, a structured variant that offers better performance than general lattices. This algebraic structure might create vulnerabilities not present in unstructured lattices. Current cryptanalysis reveals no weaknesses, though the algorithms have received substantially less scrutiny than RSA or ECC, which have been studied for decades.

SLH-DSA bases security on hash function collision resistance. Grover's algorithm provides a quadratic speedup for finding preimages, requiring twice the hash output length for equivalent post-quantum security. If collision resistance were to fail for SHA-256 or SHA-3-256, which is considered unlikely given the extensive analysis these functions have received, SLH-DSA security would be compromised. The stateless design avoids the key-management challenges of earlier hash-based signatures such as XMSS, providing conservative security in contexts where signature size is acceptable.

Real-world deployment data reveals the performance characteristics that shape migration strategies. Cloudflare reports 1.8\% of TLS 1.3 connections use post-quantum cryptography \cite{cloudflare2024pq}. Hybrid X25519+Kyber768 handshakes add 2.3 kilobytes and a median latency of 10--20 milliseconds. AWS integrated ML-KEM into Key Management Service, and Microsoft includes post-quantum options in Windows 11 and Azure \cite{aws2024pqc, microsoft2025pqc}. These deployments demonstrate feasibility but also reveal challenges in bandwidth-constrained environments and in maintaining compatibility with legacy systems.

Performance varies substantially by platform. ML-KEM-768 achieves 150 operations per second on ARM Cortex-A53 processors versus 85,000 on Intel i7-12700K and 13.3 million on NVIDIA H100 GPUs \cite{nvidia2024cupqc}. The range spanning two orders of magnitude between edge devices and high-end servers necessitates algorithm selection based on the deployment context. Energy consumption increases 15--50\% for post-quantum operations on embedded systems, reflecting differences between algorithms (ML-KEM versus SLH-DSA) and platforms (ARM versus x86). Memory requirements increase 2--5$\times$, creating additional constraints for resource-limited devices where hardware replacement costs exceed those of software updates. These constraints particularly affect IoT deployments, smart cards, and embedded systems with long operational lifecycles.

\begin{table*}[htbp]
\centering
\caption{Algorithm Performance Comparison}
\label{tab:pqc_performance}
\begin{tabular}{|l|c|c|c|c|}
\hline
\textbf{Algorithm} & \textbf{Key/Sig Size} & \textbf{Ops/sec} & \textbf{Ops/sec} & \textbf{Energy} \\
 & \textbf{(bytes)} & \textbf{(ARM A53)} & \textbf{(i7-12700K)} & \textbf{vs RSA} \\
\hline
RSA-2048 & 256 & 450 & 12,000 & 1.0x \\
ECDSA-P256 & 64 & 1,200 & 25,000 & 0.8x \\
ML-KEM-768 & 1,184/1,088 & 150 & 85,000 & 1.3x \\
ML-DSA-65 & 2,701 & 90 & 45,000 & 1.5x \\
SLH-DSA-128f & 17,088 & 12 & 3,500 & 2.1x \\
Falcon-512 & 666 & 420 & 110,000 & 1.2x \\
\hline
\end{tabular}
\end{table*}

\subsection{Artificial Intelligence in Cryptanalysis}

The literature reveals a three-way capability split in AI-based cryptographic threats that carries direct implications for defensive strategy. Specialized neural networks achieve high success rates in attacking the physical implementation layer through side-channel analysis, where the task aligns with established strengths of convolutional architectures in pattern recognition over continuous analog signals. Large language models fail comprehensively at algorithm-level cryptanalysis, achieving zero percent success on any properly implemented modern encryption algorithm, a limitation grounded in fundamental architectural constraints rather than insufficient scale or training data. Between these extremes, a third capability class has emerged in which large language models demonstrate partial effectiveness in identifying implementation-level vulnerabilities in cryptographic code, occupying a distinct threat surface from both side-channel analysis and algorithm-level attacks. This three-way bifurcation reflects the interaction between AI architectural properties and the specific characteristics of each attack surface, and understanding it is essential for accurate threat assessment and resource allocation during post-quantum migration.

\textbf{Neural Network Side-Channel Analysis.} Neural networks demonstrate significant effectiveness in extracting cryptographic keys from physical implementation characteristics. Deep learning models have been shown to extract AES keys using 570 traces from unprotected implementations \cite{wu2024deep}, compared to thousands or millions of traces required by classical differential power analysis. The reported improvement compares to specific baseline methods on specific implementations in controlled laboratory conditions rather than indicating universal success rates. Convolutional neural networks achieve high accuracy in this domain because side-channel analysis fundamentally involves pattern recognition in continuous analog signals, including power consumption, electromagnetic radiation, and timing variations. CNNs automatically learn discriminative features from these signals without manual feature engineering. Multi-modal learning combines multiple side-channel sources, achieving high success rates against protected implementations that resist single-channel attacks \cite{wu2024multimodal}. Recent work demonstrates key extraction from electromagnetic emanations at distances of 25 meters using consumer hardware \cite{cagli2024electromagnetic}, extending attack range beyond traditional side-channel distances, though signal quality degrades with distance and real-world environments introduce noise absent in laboratory settings. Hybrid CNN-Mamba architectures with residual blocks achieved a Guessing Entropy of 1 with fewer than 100 attack traces on AES-128 \cite{mamba2025sidechannel}, representing a 5--10$\times$ improvement over traditional correlation power analysis. Transfer learning enables models trained on one implementation to attack related systems with limited device access, reducing the data requirements that previously limited the practical applicability of profiled side-channel attacks, though effectiveness depends on the similarity between source and target implementations. These developments mean that attack capabilities previously limited to nation-state actors with specialized expertise are becoming more broadly accessible, though exploitation still requires domain knowledge beyond automated detection.

The threat extends to post-quantum cryptography implementations. Single-trace power analysis attacks against CRYSTALS-Kyber key generation were published at TCHES 2025 \cite{hal05455454}, demonstrating that newly standardized algorithms are vulnerable to the same class of implementation attacks that affect classical cryptography. Deep learning attacks have successfully defeated Rotating S-boxes Masking, a protection scheme specifically designed to resist conventional power analysis, achieving substantial reductions in the number of required traces compared to classical methods \cite{kuroda2021rsm}.

\textbf{Countermeasure Asymmetry.} A finding with direct implications for defensive strategy concerns the asymmetry between countermeasure effectiveness against classical and AI-enhanced attacks. First-order masking increases the number of traces required for classical attacks by a factor of $d!$, where $d$ is the masking order, but increases the requirement by only approximately 3$\times$ against neural network-based attacks \cite{primas2024masking}. Shuffling and noise injection, which provide meaningful protection against classical differential power analysis, offer limited protection against deep learning models that can learn to compensate for these perturbations. This asymmetry means that side-channel countermeasures designed for classical attack profiles provide substantially reduced protection against AI-enhanced attacks, requiring development of countermeasure strategies specifically tailored to neural network threat models. The implication for post-quantum migration is that organizations cannot assume existing countermeasure frameworks will transfer effectively to the AI-enhanced threat environment, even when those frameworks have proven effective against classical statistical attacks.

\textbf{Neural Distinguishers for Block Ciphers.} Despite comprehensive failures on production cryptography, neural networks achieve partial success against weakened cipher variants. Neural distinguishers have been shown to break 13 rounds of SIMON and 11 rounds of SPECK in related-key settings \cite{chen2024neural}, although full SIMON has 32--44 rounds depending on the key size. Neural distinguishers outperform classical differential cryptanalysis in some scenarios but require large training datasets and do not generalize to full-round ciphers. Earlier work on Speck32/64 demonstrated similar partial success \cite{gohr2019neural}, establishing that neural methods can identify statistical patterns in reduced-round ciphers but have not demonstrated capability against properly designed, full-round systems. A rigorous theoretical framework bridging learning theory and symmetric cryptanalysis introduces the conjunctive parity form (CPF) concept class and demonstrates that while a learning algorithm with sub-exponential complexity can theoretically identify certain cryptographic distinguishers, empirical analysis reveals exponential scaling costs with respect to input size and clause Hamming weight, constraining practical applicability.

\textbf{Algorithm-Level Cryptanalytic Limitations.} Large language models fail comprehensively at cryptographic reasoning tasks requiring mathematical analysis. The CipherBank benchmark, published at ACL 2025, evaluated 18 state-of-the-art models on 2,358 decryption problems spanning nine classical encryption algorithms \cite{jiang2025cipherbank}. Claude-3.5-Sonnet achieved the highest accuracy at 45.14\%, with OpenAI's o1 reasoning model at 40.59\%. DeepSeek-R1 scored 25.91\%, while most open-source alternatives fell below 10\%. Even the highest-performing models failed on the majority of polyalphabetic ciphers developed in the early modern period, and every tested model achieved 0\% success on properly implemented modern encryption algorithms including AES, RSA, and elliptic curve cryptography. These results indicate that decades of cryptographic engineering have successfully eliminated the statistical patterns that neural networks exploit in simpler ciphers. Performance varied by cipher complexity, with models achieving 60--80\% success on simple substitution ciphers such as ROT13 and Atbash but falling to near-zero accuracy on polyalphabetic ciphers such as Vigen\`{e}re, where most models scored under 2\%.

Parallel benchmarks confirmed these findings. An evaluation of 4,509 samples across nine encryption methods revealed several failure modes \cite{jiang2025cipherbank}. Performance dropped 15--20 percentage points on specialized domains (medical, technical text) compared to general literature, and token inflation emerged as a bottleneck, with ciphers expanding text representation 7--8$\times$ significantly reducing decryption success. OpenAI's o1 System Card reported 46\% accuracy on high school Capture The Flag cryptography challenges but only 13\% at professional level. Independent testing revealed instances where o1-Pro generated confident but entirely fabricated decryptions, illustrating the tendency of language models to produce plausible-sounding but incorrect outputs when confronted with tasks that exceed their reasoning capabilities. Research employing mutual information neural estimation to evaluate cryptosystem robustness in a known-plaintext setting has confirmed that properly implemented CPA-secure schemes (AES in appropriate modes) exhibit no exploitable signal, while deterministic or improperly configured schemes (DES, RSA without padding, AES in ECB mode) leak measurable mutual information.

\textbf{Theoretical Foundations.} Mechanistic interpretability research provides the architectural basis for these failures across multiple independent lines of evidence. Causal intervention studies on Llama3 and GPT-J demonstrated that large language models do not implement traditional algorithms \cite{nikankin2024arithmetic}. Instead, arithmetic reasoning emerges from collections of memorized heuristics, termed ``bags of heuristics,'' consisting of simple rules that activate for specific input patterns. Activation patching applied to discover sparse MLP circuits classified 91\% of arithmetically important neurons into distinct heuristic types, namely range heuristics (activating for specific numerical ranges), modulo heuristics (responding to remainders), and pattern heuristics (detecting digit patterns). Each neuron implements pattern matching rather than algorithmic steps.

The GSM-Symbolic study provided substantial empirical support, testing over 25 models including GPT-4o, o1, and Gemini variants \cite{mirzadeh2024gsm}. By creating symbolic variants of math problems that changed only numerical values, the study demonstrated that language models replicate reasoning steps from training data rather than performing genuine logical reasoning. Performance variance of 12--15\% across mathematically equivalent problems confirmed reliance on surface patterns rather than abstract mathematical understanding. The GSM-NoOp test, which introduced seemingly relevant but actually irrelevant information, produced substantial accuracy reductions across all tested models. This fragility to irrelevant perturbation is the opposite of what cryptanalysis requires, where a cryptanalytic tool must be invariant to surface features and responsive only to mathematical structure.

Communication complexity theory has been applied to prove that transformer architectures are mathematically incapable of composing functions when domains exceed relatively small sizes \cite{peng2024limitations}. Cryptographic algorithms fundamentally involve multi-round compositional transformations, with AES using 10--14 rounds and RSA requiring modular exponentiation over keyspaces of $2^{128}$ or larger, placing them well beyond the compositional capacity that transformers can represent. The RASP-Generalization Conjecture proposes that transformers only learn length-generalizing solutions when tasks have short RASP-L programs \cite{zhou2024algorithms}. Standard transformers exhibited little to no length generalization on arithmetic when trained from scratch, providing a theoretical explanation for cryptanalysis failures that require compositional reasoning over long sequences.

Two additional theoretical perspectives reinforce these findings. The flat loss landscape problem arises because neural networks learn by gradually reducing error along smooth gradients, whereas in cryptanalysis almost-correct decryption remains indistinguishable from random output. The loss landscape for a cryptographic task is essentially flat everywhere except at the correct key, providing no gradient signal for optimization to follow. Modern cryptographic algorithms are specifically designed to achieve computational indistinguishability, meaning that ciphertext outputs are statistically indistinguishable from random bitstrings, eliminating the correlations that machine learning models would need to exploit. Separately, work on the ``Cryptographic Wall'' concept has demonstrated that large language models are fundamentally unable to simulate the cascading, non-linear transformations inherent in cryptographic hash functions such as MD5, confirming that the same class of operations underpinning modern encryption algorithms lies beyond the computational reach of current language model architectures.

\textbf{LLM-Based Implementation Analysis.} While large language models cannot attack the mathematical algorithm layer, a distinct capability has emerged in their ability to identify implementation-level vulnerabilities in cryptographic code. This represents a third attack surface, separate from both side-channel analysis and algorithm-level cryptanalysis, in which language models analyze source code rather than ciphertext or physical emanations.

The CryptoScope framework combines Chain-of-Thought prompting with Retrieval-Augmented Generation, guided by a curated cryptographic knowledge base containing over 12,000 entries. Evaluated on a benchmark of 92 cases derived primarily from real-world CVE vulnerabilities across 11 programming languages, the framework improved performance over strong language model baselines by 11--29\% depending on the underlying model. When applied to 20 real-world cryptographic codebases, CryptoScope identified various logic-level flaws including improper ECDSA signature range checks, insecure padding in RSA, ECB-mode misuse, and weak key derivation practices, with 9 previously undisclosed flaws identified in widely used open-source cryptographic projects. Direct application of language models without domain-specific augmentation produced over 50\% false positives, but reliability improved to approximately 90\% when augmented with domain-specific knowledge retrieval and code validation techniques.

The CryptoFormalEval benchmark assesses language model capability to autonomously identify vulnerabilities in cryptographic protocols through interaction with Tamarin, a theorem prover for protocol verification \cite{cryptoformeval2024}. The benchmark tasks models with protocol formalization and proof assistance against a manually validated dataset of novel, flawed communication protocols. No model achieved perfect scores, though several frontier models demonstrated meaningful capacity for handling the intricacies of cryptographic protocol specifications. The benchmark establishes that language models can contribute to defensive cybersecurity through integration with symbolic reasoning systems but cannot autonomously replace human cryptographic expertise.

KryptoPilot represents the most recent advance in this capability class, introducing a knowledge-augmented language model agent for automated cryptographic exploitation in Capture The Flag competitions. The system integrates dynamic knowledge acquisition, structured persistent workspaces, and governed reasoning. Experimental results demonstrate that precise, executable knowledge and reasoning governance are essential for reliable exploitation, and that for high-difficulty cryptographic tasks requiring precise mathematical modeling, summary-level knowledge provides limited benefit and may mislead reasoning, necessitating access to primary research. An emerging defensive application leverages language models to accelerate hardware-software co-design for post-quantum cryptography, with language model-generated accelerators achieving up to 2.6$\times$ speedup in kernel execution time for the FALCON digital signature scheme compared to conventional HLS-based approaches.

\textbf{Cryptanalytic Methods Applied to Neural Networks.} A complementary line of research applies cryptanalytic attack methodologies to neural network models rather than using models for cryptanalysis, a distinction that reveals model vulnerabilities rather than model capabilities. Foundational work at CRYPTO 2020 pioneered treating model extraction as a cryptanalysis problem, developing attacks that exploit the piecewise linear nature of ReLU networks \cite{carlini2020cryptanalytic}. Subsequent work achieved the first successful extraction attack on production language models, extracting embedding projection layers for under \$20 USD by exploiting API design decisions made for research convenience \cite{carlini2024stealing}. Related work on training data extraction showed that language models memorize and output verbatim training data, with larger models exhibiting greater vulnerability to memorization. This line of research illustrates a directional asymmetry in that cryptanalytic methodologies prove effective against neural network models, whereas neural networks have not demonstrated capability against properly implemented cryptographic algorithms.

\textbf{Neurosymbolic Approaches.} Future progress in AI-assisted cryptanalysis is likely to require neurosymbolic architectures that combine neural pattern recognition with symbolic verification. Google DeepMind's AlphaProof couples a pre-trained Gemini model with AlphaZero reinforcement learning, achieving silver medal performance at IMO 2024 (28/42 points) \cite{deepmind2024alphageometry}. AlphaGeometry 2 solved 83\% of historical IMO geometry problems using hybrid approaches in which language models provide intuition while symbolic deduction engines ensure correctness. Both systems require formal verification frameworks and cannot operate solely through neural reasoning. DeepSeek-Prover-V2 established state-of-the-art performance at 88.9\% on miniF2F using recursive theorem proving \cite{deepseek2025prover}, though the accompanying analysis explicitly notes that verifiable rewards function effectively for mathematics and coding but that no mechanism currently exists to verify other forms of reasoning, a category that encompasses most cryptanalysis tasks where arbitrary ciphers lack mechanical verification procedures. The absence of published successes at cryptography conferences such as CRYPTO and EUROCRYPT, despite active research at machine learning conferences such as NeurIPS and ICML, suggests that the cryptography community does not yet consider AI viable for algorithm-level attacks, a professional assessment grounded in deep domain expertise and reflecting the theoretical limitations documented above.

\subsection{Quantum Key Distribution and Migration Frameworks}

China's CN-QCN represents the largest operational quantum network, spanning 10,103 kilometers with 145 backbone nodes across 17 provinces \cite{zhang2025quantum}. The network achieves secure key rates of 9.75--359.89 kbps with 99.5\% availability, demonstrating operational capability at national scale. Integration with the Micius satellite enables intercontinental quantum communication between Beijing and Vienna. However, the network requires trusted nodes every 50--150 kilometers due to photon loss and the absence of practical quantum repeaters. The designation ``quantum secure'' refers to security against interception on individual links, whereas end-to-end security requires trusting intermediate nodes, which constitutes a significant practical limitation.

Commercial deployments include Toshiba systems in Japan, ID Quantique in Geneva, and QuantumCTek, which serves thousands of Chinese users \cite{idq2024commercial}. The market is projected to grow from \$446 million in 2024 to \$2.63 billion by 2030. Technical advances include measurement-device-independent QKD, which eliminates detector vulnerabilities and achieves 537 kbps at 100 kilometers \cite{liu2024mdi}, and twin-field protocols that extend distances beyond 1,000 kilometers, albeit with increased implementation complexity.

QKD costs range from \$50,000--500,000 per node with \$10,000--50,000 per kilometer of infrastructure, limiting deployment to high-value applications. Post-quantum cryptography provides quantum resistance at software-update cost, making an economic comparison unfavorable for QKD in most deployment contexts. Physical security requirements restrict QKD to fixed installations with environmental controls, whereas post-quantum cryptography operates on any digital channel. QKD provides information-theoretic security against interception under the assumption of perfect implementation, though real devices exhibit vulnerabilities that quantum hacking attacks have exploited in practice.

Cryptographic agility has emerged as essential for managing the transition. NIST defines crypto-agility as the ability to rapidly transition between cryptographic algorithms while maintaining security and operations \cite{nist2024agility}. Organizations implementing agility frameworks achieve algorithm replacement in days rather than months, with 40--60\% reduction in future migration costs. The Global Risk Institute's Crypto Agility Maturity Model defines four levels, ranging from basic inventory to AI-driven autonomous management \cite{gri2024maturity}. Survey data indicate that only 15\% of organizations reach Level 3 (automated discovery and updates), with the majority remaining at Levels 1 and 2.

AWS migration experience demonstrates enterprise-scale implementation challenges and solutions \cite{aws2024migration}. Their systematic approach encompasses system inventory, public endpoint updates, root of trust replacement, and session authentication transition, adding 15--25\% initial infrastructure investment but reducing vulnerability windows from months to days. Microsoft's experience reveals protocol ossification affecting 15--20\% of network equipment, with certificate size increases breaking assumptions in legacy systems \cite{microsoft2025pqc}. Success factors include executive sponsorship, dedicated teams, vendor coordination, and phased rollout. Organizations that begin migration during the current period benefit from mature tools and documented practices, whereas those that defer face increasing complexity as standards evolve and compliance deadlines approach.

\subsection{Homomorphic Encryption in the Post-Quantum Era}

Fully homomorphic encryption (FHE) enables computation on encrypted data without decryption, providing a complementary privacy-preserving capability alongside post-quantum cryptography. Modern FHE schemes share mathematical foundations with lattice-based post-quantum algorithms, creating natural synergies for quantum-resistant secure computation \cite{martins2024homomorphic}. The CKKS, BGV, and BFV schemes all rely on the Ring-LWE or Module-LWE hardness assumptions that underpin ML-KEM, meaning that the security of FHE benefits directly from the cryptanalytic scrutiny applied to NIST's post-quantum standards.

The Orion framework demonstrates practical FHE for deep learning inference, achieving encrypted neural network evaluation with overhead factors of 100--1000$\times$ compared to plaintext computation \cite{orion2025fhe}. While this overhead remains prohibitive for general-purpose computing, specific applications in healthcare (privacy-preserving genomic analysis), finance (encrypted credit scoring), and government (classified data processing) justify the computational cost. Hardware acceleration through FPGA and ASIC implementations reduces overhead to 10--50$\times$ for targeted workloads, with GPU-accelerated libraries achieving further improvements.

The convergence of FHE and post-quantum cryptography creates both opportunities and challenges for the migration landscape. Organizations deploying lattice-based FHE automatically obtain quantum resistance for their encrypted computation pipelines. Conversely, the computational intensity of FHE amplifies the performance challenges already present in post-quantum algorithms. Memory requirements for FHE ciphertexts, typically 1--100 MB per encrypted value, substantially exceed even the largest post-quantum key sizes, creating distinct infrastructure demands. Organizations pursuing both privacy-preserving computation and post-quantum migration must coordinate these parallel transitions to avoid redundant infrastructure investments.

\subsection{Quantum Random Number Generation}

Quantum random number generation (QRNG) addresses a foundational requirement for cryptographic security, namely the generation of truly unpredictable random values. Classical pseudo-random number generators (PRNGs) derive randomness from deterministic algorithms seeded with environmental entropy, creating a theoretical vulnerability if the seed state is compromised. Quantum random number generators exploit fundamental quantum mechanical processes, including photon detection, vacuum fluctuations, and quantum phase noise, to produce randomness that is provably unpredictable under the laws of physics \cite{herrero2024quantum}.

Commercial QRNG devices have reached deployment maturity. ID Quantique's Quantis product line provides certified quantum randomness at rates exceeding 200 Mbps, integrated into standard PCIe and USB form factors \cite{idq2024qrng}. Samsung has embedded QRNG chips in consumer smartphones, demonstrating that quantum randomness generation is feasible at the consumer device scale. Cloud-based QRNG services from providers including AWS and IBM enable applications to access quantum randomness without dedicated hardware.

The relevance of QRNG to post-quantum cryptography is twofold. First, lattice-based algorithms such as ML-KEM and ML-DSA require high-quality randomness for key generation and signing operations. The security proofs for these algorithms assume access to a true random source, and weaknesses in the random number generator directly translate to weaknesses in the cryptographic scheme. The ML-DSA signing vulnerability discussed in Section~3.3, where single-bit randomness leakage enables a 20\% attack improvement, illustrates how randomness quality affects post-quantum security. Second, QRNG provides defense-in-depth against adversaries who might compromise classical entropy sources through supply chain attacks on hardware random number generators or side-channel extraction of PRNG states.

Several integration challenges affect current QRNG deployment. QRNG devices require calibration and self-testing to ensure that the quantum source operates correctly, as device imperfections can introduce classical correlations that reduce randomness quality. Certification standards (NIST SP 800-90B, AIS 31) provide frameworks for evaluating randomness quality, but quantum-specific certification criteria are still evolving. The cost of QRNG hardware (\$500--5,000 per device) limits deployment to high-security applications, though declining costs and chip-scale integration are expanding accessibility.

\subsection{Hybrid Protocol Architectures}

The transition from classical to post-quantum cryptography necessitates hybrid protocol designs that combine classical and post-quantum algorithms to provide security guarantees during the migration period. Hybrid approaches hedge against the possibility that newly standardized post-quantum algorithms may contain undiscovered vulnerabilities, while simultaneously protecting against quantum attacks on classical algorithms.

The IETF hybrid key exchange draft specifies the X25519+ML-KEM-768 composite key agreement for TLS 1.3 \cite{stebila2024hybrid}. The construction concatenates classical and post-quantum key shares, whereby the client sends both an X25519 ephemeral public key and an ML-KEM-768 encapsulation key in the ClientHello message. The server responds with both an X25519 public key and an ML-KEM-768 ciphertext. The shared secret is derived by concatenating both shared secrets and processing them through the TLS 1.3 key schedule, yielding $\text{SS}_{\text{hybrid}} = \text{HKDF}(\text{SS}_{X25519} \| \text{SS}_{ML\text{-}KEM})$. This construction ensures that the hybrid scheme is at least as secure as the stronger component, provided the key derivation function is secure.

Browser vendors have led hybrid deployment. Google Chrome enabled X25519+Kyber768 by default in version 124 (April 2024), immediately covering a substantial fraction of web traffic \cite{chrome2024pqc}. Mozilla Firefox followed with similar support. Cloudflare's measurements indicate that hybrid handshakes add 1,088 bytes to the ClientHello and approximately 10--20 milliseconds median latency compared to classical X25519-only handshakes. These overhead measurements proved acceptable for web traffic, though bandwidth-constrained environments such as satellite communications and IoT networks may require optimization.

Beyond TLS, hybrid approaches are being adopted across the protocol stack. Signal implemented the PQXDH (Post-Quantum Extended Diffie-Hellman) protocol, combining X25519 with ML-KEM-1024 for initial key agreement in its messaging protocol \cite{signal2024pqxdh}. Apple's iMessage adopted a similar hybrid PQ3 protocol. SSH implementations (OpenSSH 9.0+) support hybrid key exchange using the sntrup761x25519-sha512 combination. IPsec/IKEv2 hybrid proposals are under development but face additional complexity from the protocol's negotiation mechanisms.

The security analysis of hybrid constructions requires careful treatment. The hybrid assumption states that the combined scheme should be secure if either component algorithm remains unbroken. Formal proofs demonstrate this property for specific constructions under standard assumptions, but the interaction between classical and post-quantum components introduces subtleties. If an attacker can selectively downgrade connections to classical-only mode by manipulating protocol negotiation, the hybrid protection is nullified. Implementations must therefore enforce hybrid mode as mandatory rather than optional to prevent downgrade attacks. Certificate-level hybridization, involving dual signatures using both classical and post-quantum algorithms, remains an open deployment challenge, as X.509 certificate formats and PKI hierarchies were not designed for multiple signature algorithms.

\subsection{Post-Quantum Public Key Infrastructure}

The migration of public key infrastructure (PKI) to post-quantum algorithms presents challenges that extend far beyond algorithm substitution. PKI systems form the trust foundation for TLS, code signing, email security, document authentication, and identity management. The X.509 certificate ecosystem encompasses root certificate authorities, intermediate CAs, certificate transparency logs, Online Certificate Status Protocol (OCSP) responders, and Certificate Revocation Lists (CRLs). Each component must be updated in coordination to maintain the chain of trust \cite{kampanakis2024pqpki}.

Certificate size inflation represents the most immediate technical challenge. A typical TLS certificate chain contains three certificates (root CA, intermediate CA, end-entity), each carrying a public key and signature. With ECDSA-P256, the total signature and key overhead is approximately 450 bytes. Migrating to ML-DSA-65 increases this to approximately 15,000 bytes, a 33$\times$ expansion. For dual-signature hybrid certificates carrying both classical and post-quantum signatures, the overhead approximately doubles again. The CA/Browser Forum has identified several consequences of this size increase \cite{cabforum2024pqc}. TLS handshakes may exceed the initial congestion window (typically 14,600 bytes), requiring additional round trips. OCSP responses grow proportionally, increasing stapling overhead. Certificate transparency logs must accommodate larger entries, affecting log operator infrastructure.

The certificate lifecycle introduces temporal coordination challenges. Root certificates typically have 20--30 year validity periods, meaning that root certificates issued today using classical algorithms will remain in trust stores well into the post-quantum era. Replacing root certificates requires coordinated updates across all relying parties, a process that historically takes 5--10 years for universal adoption. Let's Encrypt, which issues approximately 50\% of all publicly trusted web certificates, has published post-quantum readiness assessments indicating that its automated certificate management protocol (ACME) requires protocol extensions for post-quantum certificate formats \cite{letsencrypt2024pqc}.

Code signing PKI faces distinct challenges. Software signing certificates for operating systems, drivers, and firmware updates require long-term signature validity, as a signed Windows driver must verify for the lifetime of the operating system version. Post-quantum code signing certificates must be deployed before quantum computers threaten the classical signatures protecting current software distributions. The harvest-now-decrypt-later threat extends to signed software, as an adversary who captures a signed binary today could forge a modified version once quantum computers enable signature forgery.

\subsection{Hardware Security Infrastructure}

Hardware security modules (HSMs) and Trusted Platform Modules (TPMs) provide the physical security boundary for cryptographic key material. These devices perform cryptographic operations within tamper-resistant enclosures, preventing key extraction even by privileged system administrators. The transition to post-quantum cryptography in hardware security infrastructure faces constraints distinct from software migration.

HSM vendors have begun incorporating post-quantum algorithm support. Thales Luna HSMs received firmware updates supporting ML-KEM and ML-DSA in late 2024 \cite{thales2024hsm}. Entrust nShield HSMs similarly added post-quantum capabilities. However, HSM deployments in critical infrastructure, spanning financial transaction processing, certificate authority key protection, and government classified systems, operate on 10--20 year replacement cycles. Organizations that recently deployed classical-only HSMs face a choice between firmware upgrades (where available), hardware replacement (expensive and operationally disruptive), or accepting deferred quantum protection for HSM-protected operations.

The Trusted Computing Group updated the TPM 2.0 specification to include post-quantum algorithm support \cite{tpm2025spec}. TPM chips embedded in servers, laptops, and IoT devices provide platform attestation, secure boot, and key storage. Unlike HSMs, TPMs are typically soldered to motherboards and cannot be independently replaced. The TPM 2.0 post-quantum extensions define new algorithm identifiers for ML-KEM and ML-DSA, but implementation depends on chip manufacturers releasing updated firmware. NIST's guidance for FIPS 140-3 validation of post-quantum modules \cite{nist2024cmvp} establishes the certification pathway, though the validation queue and testing requirements introduce 12--18 month delays between algorithm availability and certified module availability.

The interaction between hardware security and post-quantum algorithms creates additional performance considerations. Post-quantum algorithms require more computational resources than classical equivalents, with ML-KEM key generation using approximately 3$\times$ more cycles than ECDH and ML-DSA signing requiring approximately 5$\times$ more cycles than ECDSA on equivalent hardware. HSMs and TPMs with fixed computational budgets may experience reduced throughput after post-quantum migration. Secure enclaves (Intel SGX, ARM TrustZone) face memory constraints, as post-quantum key material and intermediate computation require larger protected memory regions.

\subsection{Post-Quantum Blockchain and Distributed Ledger Security}
\label{sec:blockchain}

Blockchain technologies present distinct post-quantum challenges due to their immutable, publicly accessible nature and decentralized governance structures \cite{allende2024quantum}. Unlike centralized systems where administrators can mandate algorithm updates, blockchain migrations require consensus among distributed participants, a process that has historically proven contentious even for minor protocol changes.

Bitcoin's quantum vulnerability derives from its use of ECDSA (secp256k1) for transaction signing. An estimated 25\% of all bitcoins reside in addresses where the public key has been exposed through address reuse, creating immediate vulnerability once CRQC becomes available \cite{stewart2024bitcoin}. Bitcoin's SHA-256 proof-of-work mechanism faces a separate quantum threat from Grover's algorithm, which would provide a quadratic speedup in mining. However, the mining difficulty adjustment mechanism would compensate for quantum speedup within approximately two weeks, making this a competitive rather than existential threat. The Bitcoin community has not yet reached consensus on a post-quantum migration strategy, with proposals ranging from a flag day hard fork to gradual migration through new address types.

Ethereum's transition strategy is more developed. The Ethereum Foundation's roadmap specifies account abstraction as the migration mechanism, enabling individual accounts to upgrade to post-quantum signature schemes without requiring a coordinated network-wide hard fork \cite{ethereum2025roadmap}. STARK-based proofs provide quantum-resistant validity proofs for Layer 2 scaling solutions, and EIP-7702 enables smart contract wallets to implement post-quantum authentication logic. Enterprise blockchain platforms such as Hyperledger Fabric and R3 Corda benefit from pluggable cryptography architectures that enable algorithm replacement through configuration changes rather than protocol modifications.

Decentralized finance (DeFi) protocols face compounded risks. Smart contracts that embed cryptographic assumptions, for example contracts using ECDSA signature verification for multi-signature wallets, cannot be updated after deployment on immutable blockchains. The total value locked in DeFi protocols exceeds \$100 billion, creating substantial financial exposure to quantum attacks on the underlying signature schemes. Post-quantum migration requires deploying new contract versions and migrating locked assets, a process complicated by governance token voting requirements and the potential for front-running during migration.

\subsection{Testing, Validation, and Interoperability}

The transition to post-quantum cryptography requires robust testing and validation methodologies that extend beyond traditional cryptographic module certification. NIST's Cryptographic Module Validation Program (CMVP) has established implementation guidance for FIPS 140-3 certification of modules incorporating post-quantum algorithms \cite{nist2024cmvp}, but the validation pipeline introduces significant delays. The average FIPS 140-3 validation currently requires 12--18 months, creating a gap between algorithm standardization and certified module availability.

Side-channel resistance testing requires specialized methodology for post-quantum implementations. The Test Vector Leakage Assessment (TVLA) framework, widely used for classical algorithms, requires adaptation for lattice-based operations where leakage patterns differ from those of AES or RSA. Formal verification of masked implementations provides higher assurance than statistical testing but demands specialized expertise \cite{primas2024masking}. The combination of formal methods and neural network-based leakage detection, using the same deep learning techniques employed by attackers, offers a pragmatic approach to implementation validation during the transition period.

Interoperability testing has revealed practical challenges in multi-vendor post-quantum deployments. Cross-vendor TLS interoperability analysis identified incompatibilities in hybrid key exchange negotiation, certificate chain validation with mixed algorithm types, and fallback behavior when post-quantum algorithms are not mutually supported \cite{sikeridis2024interop}. The Open Quantum Safe (OQS) project provides reference implementations through liboqs \cite{liboqs2024} and integration with OpenSSL \cite{openssl2024pqc}, enabling standardized testing across platforms. However, production deployments frequently deviate from reference implementations through performance optimizations or platform-specific adaptations, requiring vendor-specific interoperability testing.

Migration validation must also address regression testing for existing functionality. Post-quantum algorithm integration can introduce subtle incompatibilities. Larger certificate sizes may trigger buffer overflows in legacy parsers, extended handshake times may cause timeout failures in load balancers, and increased memory consumption may trigger out-of-memory conditions in containerized environments. Organizations require comprehensive test suites that validate not only cryptographic correctness but also system-level behavior under post-quantum configurations.
\section{Analysis} \label{sec:analysis}

\subsection{Threat Timelines and Risk Assessment}

The Global Risk Institute's 2024 survey of 47 quantum experts indicates a 14\% probability of cryptographically relevant quantum computers by 2029, 34\% by 2034, 55\% by 2039, and 79\% by 2044 \cite{mosca2024quantum}. These probabilities doubled from the 2022 assessment, which estimated a 17\% probability by 2034. The acceleration reflects concurrent progress in error correction hardware \cite{google2024willow}, revised vendor roadmaps \cite{ibm2023roadmap}, and algorithmic improvements that reduced physical qubit requirements for RSA-2048 factorization from approximately 20 million to under 1 million \cite{gidney2025quantum}. However, expert forecasts in quantum computing carry substantial uncertainty. The field has faced repeated delays over the past two decades, and unforeseen engineering challenges in areas such as interconnect scaling, cryogenic infrastructure, and control electronics could further extend timelines. The probabilities represent consensus estimates based on current progress rather than certainties about future capabilities.

Mosca's risk equation ($X + Y > Z$) provides a decision framework in which the data sensitivity period ($X$) plus the migration time ($Y$) must not exceed the time to quantum threat ($Z$) \cite{mosca2024quantum}. With a 34\% probability of CRQC by 2034, organizations with data sensitivity periods of 10 years and migration timelines of 5--10 years satisfy the inequality under most quantum timeline estimates. Financial services organizations that protect transaction records, healthcare organizations that maintain patient data with 50+ year sensitivity periods, and government agencies that secure classified information with 25--75+ year sensitivity periods each produce $X + Y$ sums that substantially exceed conservative estimates of $Z$. Even organizations with shorter data sensitivity periods must account for extended migration timelines, as the combined period often exceeds the lower bound of quantum threat estimates.

The AI threat landscape requires analysis through the three-way capability split documented in Section~3. Neural network side-channel attacks represent the most immediate threat, with capabilities that are operational today rather than contingent on future hardware development. Current neural cryptanalysis extracts keys from unprotected and weakly protected implementations, though the practical threat to full-round ciphers with robust side-channel countermeasures remains limited. The open question of whether theoretical lower bounds exist on the number of traces required for neural key recovery determines whether countermeasures can provide lasting protection or whether increasing model capacity will progressively erode defensive effectiveness. Large language model-based code analysis tools such as CryptoScope represent an emerging capability that can identify implementation vulnerabilities in cryptographic libraries with approximately 90\% reliability when augmented with domain-specific knowledge, though direct application without augmentation produces false positive rates exceeding 50\%. Algorithm-level cryptanalysis by large language models remains at zero percent success on any properly implemented modern encryption algorithm, a limitation grounded in fundamental architectural constraints documented in Section~3.

Transfer learning reduces data requirements for side-channel attacks, enabling attacks on related implementations with limited device access, though effectiveness depends on the similarity between source and target implementations. Automated vulnerability discovery accelerates the identification of potential leakage points, though exploitation still requires domain expertise beyond detection alone. The countermeasure asymmetry documented in Section~3, in which first-order masking increases classical attack traces by $d!$ but only approximately 3$\times$ against neural networks \cite{primas2024masking}, means that existing defensive frameworks provide substantially reduced protection against AI-enhanced attacks compared to their effectiveness against classical statistical methods.

The combination of quantum and AI capabilities creates compound threat models that existing security proofs do not address. Current security analyses assume either classical or quantum adversaries but do not account for adversaries who combine polynomial-time quantum algorithms with neural network-based side-channel analysis. Quantum computers could in principle accelerate neural network training, enabling larger models to analyze more complex side-channel patterns. Quantum machine learning remains experimental, but the potential for future convergence warrants attention. Formal security analysis under hybrid adversary models, in which attackers have access to both quantum computational advantages and AI-enhanced implementation attacks, remains an open research problem that creates uncertainty about actual security margins for deployed systems.

\subsection{Algorithm Security and Implementation Challenges}

Lattice-based algorithms (ML-KEM, ML-DSA) derive their security from the computational hardness of the Shortest Vector Problem (SVP) and the Learning with Errors (LWE) problem. Worst-case-to-average-case reductions provide a theoretical foundation in that breaking average-case instances implies solving worst-case lattice problems, which are believed to be computationally intractable. This reduction structure differs from RSA, which relies on average-case hardness without worst-case reduction. However, ML-KEM and ML-DSA utilize module lattices, a structured variant that offers better performance than general lattices but might create vulnerabilities not present in unstructured lattice problems. Current cryptanalysis reveals no weaknesses, though the algorithms have received substantially less scrutiny than RSA or ECC, which have withstood decades of intensive cryptanalytic effort.

Hash-based signatures (SLH-DSA) rely solely on collision-resistant hash functions, meaning that breaking SLH-DSA requires breaking the underlying hash function. SHA-256 and SHA-3-256 have received extensive analysis over decades of deployment. Grover's algorithm halves effective security but does not break properly sized hashes, and the stateless design of SLH-DSA avoids the key-management challenges of earlier hash-based schemes. The security comparison favors hash-based approaches for conservative deployments where signature sizes of 17--50 KB are acceptable. Lattice-based approaches offer substantially better performance characteristics, though they rest on mathematical foundations that are theoretically sound but less extensively tested in practice.

Implementation vulnerabilities manifest differently across algorithm types, and the three-way AI capability split documented in Section~3 creates distinct threat profiles for each. Lattice-based algorithms employ rejection sampling, which introduces timing variations that may leak information about secret values. ML-DSA signing involves sampling from discrete Gaussian distributions, where power consumption and electromagnetic radiation during sampling can leak secret key information through analog side channels that neural networks are well-suited to exploit. Hash-based signatures have simpler operations consisting primarily of hash function evaluations and XOR operations, reducing the number of side-channel leakage sources and correspondingly reducing the attack surface available to neural network-based analysis. The practical implication is that lattice-based algorithms, while offering superior performance, require more extensive implementation hardening than hash-based alternatives to achieve equivalent resistance to AI-enhanced side-channel attacks.

Table~\ref{tab:countermeasure_taxonomy} presents a taxonomy of side-channel countermeasures and their effectiveness against classical statistical attacks versus AI-enhanced neural network attacks. The table reveals that countermeasures providing strong protection against classical differential power analysis may offer substantially reduced effectiveness against deep learning attacks. Masking, which distributes sensitive values across multiple random shares, increases the number of traces required for classical attacks by a factor of $d!$ for $d$-th order masking. However, neural networks can learn higher-order statistical relationships, reducing the effective protection to approximately 3$\times$ the unmasked trace count for first-order masking \cite{primas2024masking}. Only blinding and threshold implementations maintain high effectiveness against both classical and AI-enhanced adversaries, though these approaches carry higher computational overhead and are not applicable to all algorithm types. This asymmetry between classical and AI-enhanced attack resistance represents a central consideration for implementation hardening strategies during post-quantum migration.

\begin{table*}[htbp]
\centering
\caption{Side-Channel Countermeasure Taxonomy: Effectiveness Against Classical vs. AI-Enhanced Attacks}
\label{tab:countermeasure_taxonomy}
\begin{tabular}{|l|p{3.5cm}|c|c|c|l|}
\hline
\textbf{Countermeasure} & \textbf{Mechanism} & \textbf{Classical} & \textbf{AI-Enhanced} & \textbf{Overhead} & \textbf{Applicable} \\
 & & \textbf{Protection} & \textbf{Protection} & & \textbf{Algorithms} \\
\hline
Constant-time code & Eliminates data-dependent branches and memory access patterns & High & Moderate & 5--20\% & All \\
\hline
Boolean masking & Splits sensitive values into random shares using XOR & Very High & Moderate & 2--10$\times$ & AES, ML-KEM \\
\hline
Arithmetic masking & Splits values using modular addition for arithmetic operations & Very High & Moderate & 3--15$\times$ & ML-DSA, FN-DSA \\
\hline
Shuffling & Randomizes order of independent operations & High & Low--Moderate & 1.5--3$\times$ & AES S-box \\
\hline
Noise injection & Adds random operations to mask signal & Moderate & Low & 1.2--2$\times$ & All \\
\hline
Hiding (amplitude) & Equalizes power consumption across operations & High & Moderate & 2--5$\times$ & All (hardware) \\
\hline
Blinding & Randomizes inputs/outputs of operations & Very High & High & 1.5--3$\times$ & RSA, ECC \\
\hline
Threshold impl. & Provably secure sharing with $d+1$ shares for $d$-th order & Very High & High & $(d+1)\times$ & AES, small ops \\
\hline
\end{tabular}
\end{table*}

Performance variations across deployment platforms create additional challenges for algorithm selection and migration planning. ML-KEM-768 achieves 150 operations per second on ARM Cortex-A53 processors versus 85,000 on Intel i7-12700K \cite{nvidia2024cupqc}, a range spanning two orders of magnitude that necessitates platform-specific algorithm selection rather than universal deployment decisions. Memory requirements increase 2--5$\times$ compared to classical equivalents, creating constraints for embedded systems and IoT devices where fixed memory cannot accommodate increases without hardware replacement. Energy consumption increases of 15--50\% affect battery-powered devices and data center cooling, with the range reflecting variation across algorithms (ML-KEM versus SLH-DSA) and platforms (ARM versus x86). Certificate and signature size increases of 10--30$\times$ challenge bandwidth-constrained environments and create multiplicative effects in complex PKI hierarchies where certificate chains contain multiple signatures and public keys. These deployment constraints are not uniformly distributed across sectors, and the binding constraint for migration feasibility differs by deployment context in ways that parallel the sector-specific analysis of Mosca's equation parameters.

\subsection{Deployment Status and Adoption Metrics}

Table~\ref{tab:deployment_metrics} consolidates current deployment metrics across the post-quantum migration landscape, providing a snapshot of progress as of mid-2025. The data reveal a consistent pattern in which client-side adoption substantially outpaces server-side deployment, key exchange migration leads digital signature migration by years, and software updates proceed faster than hardware infrastructure changes. These asymmetries reflect fundamental differences in deployment complexity, update mechanisms, and stakeholder coordination requirements.

\begin{table*}[htbp]
\centering
\caption{Post-Quantum Deployment Metrics (Mid-2025)}
\label{tab:deployment_metrics}
\begin{tabular}{|l|l|c|l|}
\hline
\textbf{Category} & \textbf{Metric} & \textbf{Value} & \textbf{Source/Notes} \\
\hline
\multicolumn{4}{|c|}{\cellcolor{gray!15}\textbf{TLS Key Exchange}} \\
\hline
Client support & Browsers supporting hybrid PQ KE & $>$95\% & Chrome 124+, Firefox, Edge \\
Client traffic & Web traffic using hybrid PQ KE & 52\% & Driven by browser defaults \\
Server support & Servers supporting hybrid PQ KE & 3.7\% & Cloudflare measurement \\
PQ-only TLS & Connections using PQ-only (no hybrid) & $<$0.1\% & Negligible adoption \\
\hline
\multicolumn{4}{|c|}{\cellcolor{gray!15}\textbf{Digital Signatures}} \\
\hline
Code signing & PQ code signing certificates issued & $<$100 & Pilot programs only \\
Web PKI & PQ certificates in Certificate Transparency & $\sim$0\% & No production issuance \\
Firmware signing & Vendors using PQ firmware signatures & $<$5 & Early adopters only \\
\hline
\multicolumn{4}{|c|}{\cellcolor{gray!15}\textbf{Infrastructure}} \\
\hline
Cloud providers & Major clouds with PQ KMS support & 3 & AWS, Azure, GCP \\
HSM vendors & HSMs with PQ algorithm support & 4 & Thales, Entrust, Utimaco, Marvell \\
FIPS 140-3 & PQ modules with FIPS validation & 0 & In validation queue \\
\hline
\multicolumn{4}{|c|}{\cellcolor{gray!15}\textbf{Messaging and Applications}} \\
\hline
Messaging & Apps with PQ key exchange & 3 & Signal, iMessage, WhatsApp \\
SSH & OpenSSH versions with PQ support & 9.0+ & sntrup761x25519 default \\
VPN & VPN products with PQ support & $<$10 & WireGuard, some commercial \\
\hline
\multicolumn{4}{|c|}{\cellcolor{gray!15}\textbf{Organizational Readiness}} \\
\hline
Crypto inventory & Organizations with complete inventory & $\sim$20\% & Most at Level 1--2 maturity \\
Crypto agility & Organizations at Level 3+ maturity & 15\% & GRI Maturity Model \\
Migration plans & Organizations with formal PQ plans & $\sim$30\% & Concentrated in regulated sectors \\
\hline
\end{tabular}
\end{table*}

The deployment data reveal several structural observations. The 52\% client-side versus 3.7\% server-side adoption gap, representing a 14$\times$ difference, demonstrates that browser vendor defaults can drive rapid client deployment, but server operators require active decision-making and infrastructure investment to achieve comparable adoption. Digital signature migration lags key exchange by years because signatures affect more system components, spanning PKI, certificate management, code signing, and document authentication, and require coordinated updates across trust hierarchies rather than unilateral deployment decisions. The absence of FIPS 140-3 validated post-quantum modules creates a compliance barrier for regulated organizations that mandate validated cryptography, introducing a gap between algorithm availability and deployable certified implementations that the 12--18 month validation timeline will not close before regulatory deadlines in several jurisdictions.

\begin{keyfinding}[title={\ding{72} Deployment Asymmetry}]
Post-quantum migration reveals a \textbf{14$\times$ gap} between 52\% client-side adoption (driven by browser defaults) and only 3.7\% server-side support. Digital signature migration remains at \textbf{$\sim$0\%}, and zero FIPS 140-3 validated post-quantum modules exist. The ecosystem is partially protected at the key exchange layer while authentication and signing remain fully exposed to future quantum attack.
\end{keyfinding}

\subsection{Risk Assessment Decision Framework}

The convergent threat landscape requires organizations to assess quantum and AI risks simultaneously rather than independently. The following framework synthesizes Mosca's risk equation with the deployment metrics and countermeasure analysis presented above to provide an integrated risk assessment across four dimensions.

\textbf{Dimension 1: Data Exposure Window.} Calculate $X + Y$ where $X$ is the data sensitivity period and $Y$ is the estimated migration time for the relevant sector. If $X + Y > 10$ years, which represents the conservative lower bound for quantum threat based on the 14\% probability of CRQC by 2029, immediate migration planning is warranted. Healthcare organizations ($X + Y \geq 60$ years), government agencies ($X + Y \geq 35$ years), and energy infrastructure operators ($X + Y \geq 35$ years) face the highest exposure under this framework.

\textbf{Dimension 2: Implementation Attack Surface.} Assess whether cryptographic operations occur on hardware susceptible to side-channel analysis. Organizations deploying cryptography on shared cloud infrastructure, embedded devices, or physically accessible hardware face higher exposure to AI-enhanced side-channel attacks. The countermeasure taxonomy (Table~\ref{tab:countermeasure_taxonomy}) maps defensive options to specific attack vectors and quantifies the asymmetry between classical and AI-enhanced protection levels for each countermeasure type.

\textbf{Dimension 3: Regulatory Compliance Pressure.} Map applicable regulatory requirements against current deployment status from Table~\ref{tab:deployment_metrics}. Organizations subject to CNSA 2.0 face 2025--2030 deadlines for quantum-resistant algorithm deployment \cite{nsa2024mandate}. DORA-regulated financial institutions require quantum risk assessments by 2025. The gap between regulatory expectations and the current absence of FIPS 140-3 validated post-quantum modules creates compliance risk that is independent of technical migration progress.

\textbf{Dimension 4: Cryptographic Agility Maturity.} Assess current agility level against the GRI maturity model \cite{gri2024maturity}. Organizations at Level 1--2, which represents 85\% of the surveyed population, face 2--5$\times$ higher migration costs and longer timelines than Level 3+ organizations. Investment in agility infrastructure provides compounding returns by reducing the cost and risk of both current and future algorithm transitions.

Organizations scoring high exposure across multiple dimensions should treat post-quantum migration as a strategic priority with executive sponsorship and dedicated resources. Organizations with lower exposure across all dimensions may adopt a phased approach, beginning with cryptographic inventory and agility investment while monitoring threat timeline developments.

\subsection{Research Gaps}

\begin{table*}[htbp]
\centering
\caption{Critical Research Gaps}
\label{tab:research_gaps}
\begin{tabular}{|p{3cm}|p{7cm}|p{3cm}|}
\hline
\textbf{Category} & \textbf{Gap} & \textbf{Priority} \\
\hline
Quantum & Hybrid quantum-classical attack analysis & Critical \\
 & Parameter security under improved algorithms & High \\
 & NISQ device cryptanalytic capability & Medium \\
\hline
AI & Lower bounds on trace requirements & Critical \\
 & Transfer learning limitations & High \\
 & Adversarial attacks on defensive AI & High \\
\hline
Implementation & Side-channel resistance verification & Critical \\
 & Formal methods for PQC implementations & High \\
 & Hardware acceleration for constrained devices & Medium \\
\hline
Integration & Hybrid protocol security proofs & Critical \\
 & Agility framework standardization & High \\
 & Migration tool maturity & Medium \\
\hline
\end{tabular}
\end{table*}

Hybrid quantum-classical attack models represent the most significant gap in current threat analysis. Existing security proofs assume either classical or quantum adversaries but do not account for adversaries who combine both paradigms. Understanding security under hybrid threat models is essential for confident parameter selection and for determining whether current security margins are sufficient against the full range of plausible future adversaries.

The question of whether theoretical lower bounds exist on the number of side-channel traces required for neural key recovery determines the long-term viability of countermeasure-based defense. If such bounds exist, side-channel countermeasures can provide guaranteed protection by pushing required trace counts below practical collection thresholds. If no such bounds exist, increasing model capacity will progressively erode defensive effectiveness regardless of countermeasure investment, and physical isolation of cryptographic operations becomes the only reliable long-term defense.

\begin{researchgap}
The most consequential open question for implementation security is whether theoretical lower bounds exist on the number of side-channel traces required for neural key recovery. If bounds exist, countermeasures can guarantee security below the collection threshold. If not, increasing model capacity will progressively erode defensive effectiveness, and physical isolation becomes the only reliable long-term defense.
\end{researchgap}

Side-channel resistance verification requires formal methods capable of proving that implementations do not leak information through analog side channels. Current testing-based approaches identify vulnerabilities but cannot prove their absence. Formal verification of post-quantum implementations would provide higher security assurance, though extending existing verification frameworks to cover the complex operations characteristic of lattice-based algorithms (rejection sampling, polynomial multiplication, discrete Gaussian sampling) remains technically challenging. The CryptoFormalEval benchmark \cite{cryptoformeval2024} provides initial evaluation criteria for language model-assisted formal verification, and tools such as Lean Copilot \cite{leancopilot2024} demonstrate that AI can assist with formal theorem proving in adjacent mathematical domains. Extending these capabilities to post-quantum implementation verification represents a high-impact research opportunity.

\section{Conclusion} \label{sec:conclusion}

Quantum computing and artificial intelligence create convergent threats to cryptographic systems that target fundamentally distinct layers of the security stack. Expert consensus indicates a 34\% probability of cryptographically relevant quantum computers emerging by 2034, while AI-enhanced side-channel analysis already demonstrates practical key recovery against protected implementations using single power traces. These threats operate independently, meaning that post-quantum algorithm deployment addresses quantum vulnerabilities but provides no protection against implementation attacks, while side-channel hardening mitigates AI-enhanced physical attacks but leaves mathematical vulnerabilities unaddressed. NIST standardization of ML-KEM, ML-DSA, and SLH-DSA provides a viable migration path, with real-world deployments demonstrating feasibility. However, deployment remains asymmetric, with client-side hybrid key exchange reaching 52\% of web traffic through browser defaults while server-side adoption remains at 3.7\% and digital signature migration approaches zero. The countermeasure asymmetry documented in this work, in which classical side-channel defenses provide substantially reduced protection against neural network-based attacks, compounds the challenge by requiring organizations to develop AI-aware defensive strategies alongside algorithm migration. No single cryptographic approach addresses both quantum and AI threats comprehensively, and defense-in-depth combining post-quantum algorithms, implementation hardening, cryptographic agility, and hardware security boundaries provides substantially better protection than any individual technique. Given that migration typically requires 5--10 years for enterprises and 10--20 years for critical infrastructure, organizations with long-term data confidentiality requirements that satisfy Mosca's inequality ($X + Y > Z$) face binding urgency under current quantum threat probability estimates.

Several open research problems constrain current confidence in security margins. Hybrid quantum-classical threat models that account for adversaries combining polynomial-time quantum algorithms with AI-enhanced side-channel analysis do not yet exist, leaving uncertainty about whether defenses designed against each threat class individually compose securely when both are present. The question of whether theoretical lower bounds exist on the number of side-channel traces required for neural key recovery determines the long-term viability of countermeasure-based defense. Formal verification methods capable of proving that post-quantum implementations are free from side-channel vulnerabilities remain technically challenging for lattice-based operations, though initial work on AI-assisted formal verification suggests a plausible path forward. Standardization of cryptographic agility frameworks and optimization of post-quantum algorithms for constrained embedded environments represent additional priorities that will shape the feasibility of migration across sectors with severe computational limitations. These gaps create uncertainty that conservative parameter selection and defense-in-depth partially mitigate, but resolving them is necessary to establish the security assurances required for confident post-quantum deployment at scale.

\bibliographystyle{IEEEtran}

\bibliography{Ref}
\end{document}